\documentclass[]{aastex631}

\usepackage{longtable}
\usepackage{booktabs}
\usepackage{comment}

\begin{document}

\title{Time-Integrated Southern-Sky Neutrino Source Searches with 10 Years of IceCube Starting-Track Events at Energies Down to 1 TeV}

\affiliation{III. Physikalisches Institut, RWTH Aachen University, D-52056 Aachen, Germany}
\affiliation{Department of Physics, University of Adelaide, Adelaide, 5005, Australia}
\affiliation{Dept. of Physics and Astronomy, University of Alaska Anchorage, 3211 Providence Dr., Anchorage, AK 99508, USA}
\affiliation{Dept. of Physics, University of Texas at Arlington, 502 Yates St., Science Hall Rm 108, Box 19059, Arlington, TX 76019, USA}
\affiliation{School of Physics and Center for Relativistic Astrophysics, Georgia Institute of Technology, Atlanta, GA 30332, USA}
\affiliation{Dept. of Physics, Southern University, Baton Rouge, LA 70813, USA}
\affiliation{Dept. of Physics, University of California, Berkeley, CA 94720, USA}
\affiliation{Lawrence Berkeley National Laboratory, Berkeley, CA 94720, USA}
\affiliation{Institut f{\"u}r Physik, Humboldt-Universit{\"a}t zu Berlin, D-12489 Berlin, Germany}
\affiliation{Fakult{\"a}t f{\"u}r Physik {\&} Astronomie, Ruhr-Universit{\"a}t Bochum, D-44780 Bochum, Germany}
\affiliation{Universit{\'e} Libre de Bruxelles, Science Faculty CP230, B-1050 Brussels, Belgium}
\affiliation{Vrije Universiteit Brussel (VUB), Dienst ELEM, B-1050 Brussels, Belgium}
\affiliation{Dept. of Physics, Simon Fraser University, Burnaby, BC V5A 1S6, Canada}
\affiliation{Department of Physics and Laboratory for Particle Physics and Cosmology, Harvard University, Cambridge, MA 02138, USA}
\affiliation{Dept. of Physics, Massachusetts Institute of Technology, Cambridge, MA 02139, USA}
\affiliation{Dept. of Physics and The International Center for Hadron Astrophysics, Chiba University, Chiba 263-8522, Japan}
\affiliation{Department of Physics, Loyola University Chicago, Chicago, IL 60660, USA}
\affiliation{Dept. of Physics and Astronomy, University of Canterbury, Private Bag 4800, Christchurch, New Zealand}
\affiliation{Dept. of Physics, University of Maryland, College Park, MD 20742, USA}
\affiliation{Dept. of Astronomy, Ohio State University, Columbus, OH 43210, USA}
\affiliation{Dept. of Physics and Center for Cosmology and Astro-Particle Physics, Ohio State University, Columbus, OH 43210, USA}
\affiliation{Niels Bohr Institute, University of Copenhagen, DK-2100 Copenhagen, Denmark}
\affiliation{Dept. of Physics, TU Dortmund University, D-44221 Dortmund, Germany}
\affiliation{Dept. of Physics and Astronomy, Michigan State University, East Lansing, MI 48824, USA}
\affiliation{Dept. of Physics, University of Alberta, Edmonton, Alberta, T6G 2E1, Canada}
\affiliation{Erlangen Centre for Astroparticle Physics, Friedrich-Alexander-Universit{\"a}t Erlangen-N{\"u}rnberg, D-91058 Erlangen, Germany}
\affiliation{Physik-department, Technische Universit{\"a}t M{\"u}nchen, D-85748 Garching, Germany}
\affiliation{D{\'e}partement de physique nucl{\'e}aire et corpusculaire, Universit{\'e} de Gen{\`e}ve, CH-1211 Gen{\`e}ve, Switzerland}
\affiliation{Dept. of Physics and Astronomy, University of Gent, B-9000 Gent, Belgium}
\affiliation{Dept. of Physics and Astronomy, University of California, Irvine, CA 92697, USA}
\affiliation{Karlsruhe Institute of Technology, Institute for Astroparticle Physics, D-76021 Karlsruhe, Germany}
\affiliation{Karlsruhe Institute of Technology, Institute of Experimental Particle Physics, D-76021 Karlsruhe, Germany}
\affiliation{Dept. of Physics, Engineering Physics, and Astronomy, Queen's University, Kingston, ON K7L 3N6, Canada}
\affiliation{Department of Physics {\&} Astronomy, University of Nevada, Las Vegas, NV 89154, USA}
\affiliation{Nevada Center for Astrophysics, University of Nevada, Las Vegas, NV 89154, USA}
\affiliation{Dept. of Physics and Astronomy, University of Kansas, Lawrence, KS 66045, USA}
\affiliation{Centre for Cosmology, Particle Physics and Phenomenology - CP3, Universit{\'e} catholique de Louvain, Louvain-la-Neuve, Belgium}
\affiliation{Department of Physics, Mercer University, Macon, GA 31207-0001, USA}
\affiliation{Dept. of Astronomy, University of Wisconsin{\textemdash}Madison, Madison, WI 53706, USA}
\affiliation{Dept. of Physics and Wisconsin IceCube Particle Astrophysics Center, University of Wisconsin{\textemdash}Madison, Madison, WI 53706, USA}
\affiliation{Institute of Physics, University of Mainz, Staudinger Weg 7, D-55099 Mainz, Germany}
\affiliation{Department of Physics, Marquette University, Milwaukee, WI 53201, USA}
\affiliation{Institut f{\"u}r Kernphysik, Universit{\"a}t M{\"u}nster, D-48149 M{\"u}nster, Germany}
\affiliation{Bartol Research Institute and Dept. of Physics and Astronomy, University of Delaware, Newark, DE 19716, USA}
\affiliation{Dept. of Physics, Yale University, New Haven, CT 06520, USA}
\affiliation{Columbia Astrophysics and Nevis Laboratories, Columbia University, New York, NY 10027, USA}
\affiliation{Dept. of Physics, University of Oxford, Parks Road, Oxford OX1 3PU, United Kingdom}
\affiliation{Dipartimento di Fisica e Astronomia Galileo Galilei, Universit{\`a} Degli Studi di Padova, I-35122 Padova PD, Italy}
\affiliation{Dept. of Physics, Drexel University, 3141 Chestnut Street, Philadelphia, PA 19104, USA}
\affiliation{Physics Department, South Dakota School of Mines and Technology, Rapid City, SD 57701, USA}
\affiliation{Dept. of Physics, University of Wisconsin, River Falls, WI 54022, USA}
\affiliation{Dept. of Physics and Astronomy, University of Rochester, Rochester, NY 14627, USA}
\affiliation{Department of Physics and Astronomy, University of Utah, Salt Lake City, UT 84112, USA}
\affiliation{Dept. of Physics, Chung-Ang University, Seoul 06974, Republic of Korea}
\affiliation{Oskar Klein Centre and Dept. of Physics, Stockholm University, SE-10691 Stockholm, Sweden}
\affiliation{Dept. of Physics and Astronomy, Stony Brook University, Stony Brook, NY 11794-3800, USA}
\affiliation{Dept. of Physics, Sungkyunkwan University, Suwon 16419, Republic of Korea}
\affiliation{Institute of Basic Science, Sungkyunkwan University, Suwon 16419, Republic of Korea}
\affiliation{Institute of Physics, Academia Sinica, Taipei, 11529, Taiwan}
\affiliation{Dept. of Physics and Astronomy, University of Alabama, Tuscaloosa, AL 35487, USA}
\affiliation{Dept. of Astronomy and Astrophysics, Pennsylvania State University, University Park, PA 16802, USA}
\affiliation{Dept. of Physics, Pennsylvania State University, University Park, PA 16802, USA}
\affiliation{Dept. of Physics and Astronomy, Uppsala University, Box 516, SE-75120 Uppsala, Sweden}
\affiliation{Dept. of Physics, University of Wuppertal, D-42119 Wuppertal, Germany}
\affiliation{Deutsches Elektronen-Synchrotron DESY, Platanenallee 6, D-15738 Zeuthen, Germany}

\author[0000-0001-6141-4205]{R. Abbasi}
\affiliation{Department of Physics, Loyola University Chicago, Chicago, IL 60660, USA}

\author[0000-0001-8952-588X]{M. Ackermann}
\affiliation{Deutsches Elektronen-Synchrotron DESY, Platanenallee 6, D-15738 Zeuthen, Germany}

\author{J. Adams}
\affiliation{Dept. of Physics and Astronomy, University of Canterbury, Private Bag 4800, Christchurch, New Zealand}

\author[0000-0002-9714-8866]{S. K. Agarwalla}
\altaffiliation{also at Institute of Physics, Sachivalaya Marg, Sainik School Post, Bhubaneswar 751005, India}
\affiliation{Dept. of Physics and Wisconsin IceCube Particle Astrophysics Center, University of Wisconsin{\textemdash}Madison, Madison, WI 53706, USA}

\author[0000-0003-2252-9514]{J. A. Aguilar}
\affiliation{Universit{\'e} Libre de Bruxelles, Science Faculty CP230, B-1050 Brussels, Belgium}

\author[0000-0003-0709-5631]{M. Ahlers}
\affiliation{Niels Bohr Institute, University of Copenhagen, DK-2100 Copenhagen, Denmark}

\author[0000-0002-9534-9189]{J.M. Alameddine}
\affiliation{Dept. of Physics, TU Dortmund University, D-44221 Dortmund, Germany}

\author{N. M. Amin}
\affiliation{Bartol Research Institute and Dept. of Physics and Astronomy, University of Delaware, Newark, DE 19716, USA}

\author[0000-0001-9394-0007]{K. Andeen}
\affiliation{Department of Physics, Marquette University, Milwaukee, WI 53201, USA}

\author[0000-0003-4186-4182]{C. Arg{\"u}elles}
\affiliation{Department of Physics and Laboratory for Particle Physics and Cosmology, Harvard University, Cambridge, MA 02138, USA}

% ...

\date{\today}

\collaboration{427}{IceCube Collaboration}

\begin{abstract}
In the IceCube Neutrino Observatory, a signal of astrophysical neutrinos is obscured by backgrounds from atmospheric neutrinos and muons produced in cosmic-ray interactions.
IceCube event selections used to isolate the astrophysical neutrino signal often focus on t/he morphology of the light patterns recorded by the detector.
The analyses presented here use the new IceCube Enhanced Starting Track Event Selection (ESTES), which identifies events likely generated by muon neutrino interactions within the detector geometry, focusing on neutrino energies of 1-500 TeV with a median angular resolution of 1.4°.
Selecting for starting track events filters out not only the atmospheric-muon background, but also the atmospheric-neutrino background in the southern sky.
This improves IceCube's muon neutrino sensitivity to southern-sky neutrino sources, especially for Galactic sources that are not expected to produce a substantial flux of neutrinos above 100 TeV.
In this work, the ESTES sample was applied for the first time to searches for astrophysical sources of neutrinos, including a search for diffuse neutrino emission from the Galactic plane. 
No significant excesses were identified from any of the analyses; however, constraining limits are set on the hadronic emission from TeV gamma-ray Galactic plane objects and models of the diffuse Galactic plane neutrino flux.
\end{abstract}



\section{Introduction} \label{sec:intro}
A main aim of the IceCube Neutrino Observatory is to locate the origins of astrophysical neutrinos in our universe. 
Neutrino emission from astrophysical sources gives insight into the production mechanisms of high-energy cosmic rays in the source environment.
In 2013, IceCube announced its discovery of a flux of astrophysical neutrinos observed at Earth; however, the arrival direction of these astrophysical candidate events was found to be consistent with an isotropic flux~\citep{IceCube:2013HESEDiscovery}.
In the decade since this discovery, two active galactic nuclei (AGN) have been identified by IceCube as likely cosmic-ray accelerators: TXS 0506+056 and NGC 1068~\citep{IceCube:TXSMMPaper, IceCube:TXSNeutFlare, IceCube:NGC1068Sci}.
Recently, IceCube published evidence for neutrinos from the Galactic plane~\citep{IceCube:DNNCSci}.
Each of these discoveries used different selections of IceCube events curated to take advantage of the various morphological structures that are created in the light produced by neutrinos interacting in or around the IceCube detector, which instruments a cubic kilometer of deep Antarctic ice.
Here, we present the results of the neutrino source searches from a new IceCube event selection: the Enhanced Starting Track Event Selection (ESTES)~\citep{IceCube:ESTESPRD}.
This novel selection improves the IceCube sensitivity to southern-sky neutrino sources by finding astrophysical neutrino candidate events previously missed in other IceCube data selections.

\begin{figure}[tb!]
    \centering
    \includegraphics[width=0.65\textwidth]{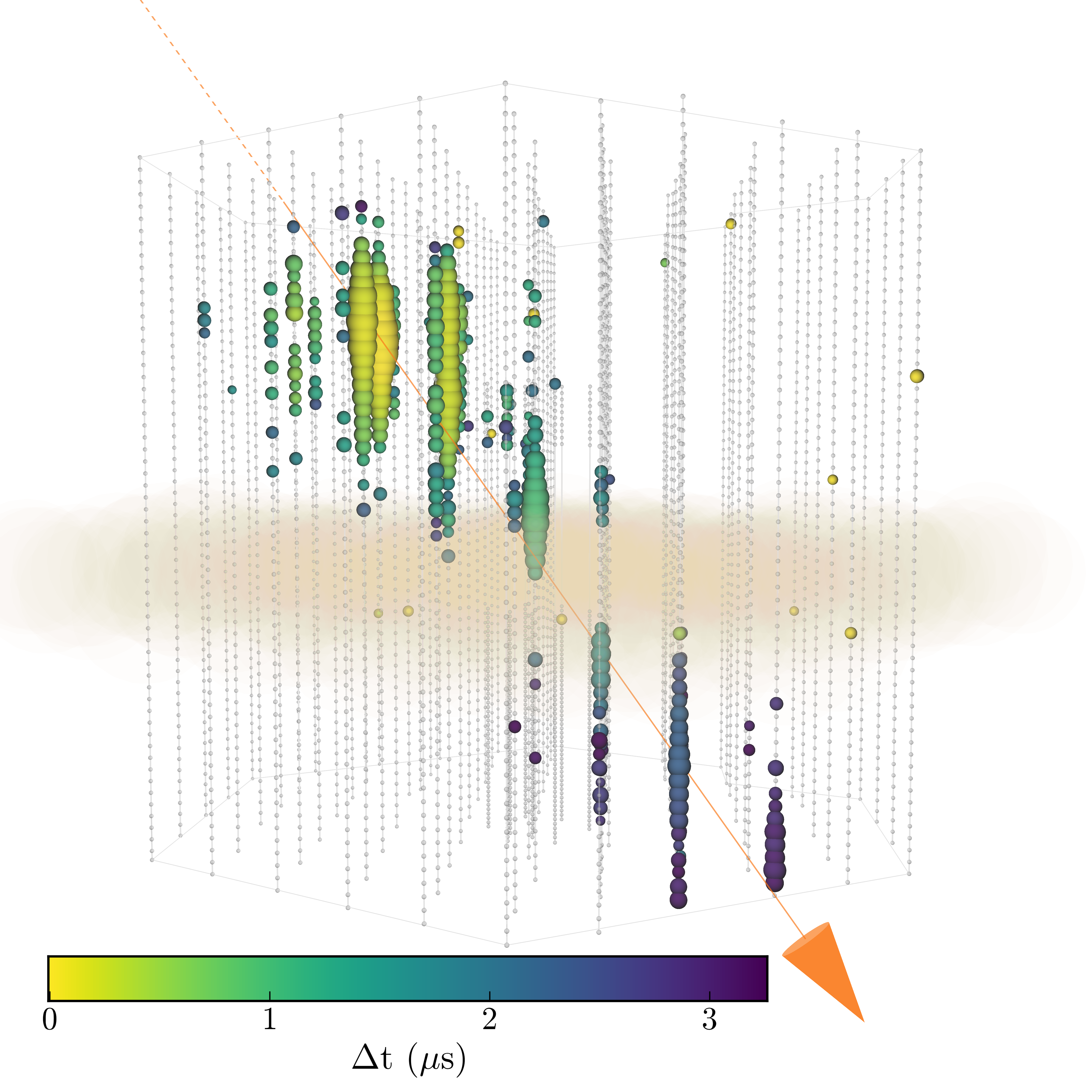}
    \caption{Visualization of a starting-track event from the final level of the enhanced starting-track event selection (ESTES) data.
    The strings (grey lines) and optical modules (spheres) depict the IceCube detector geometry buried in the glacial ice at the South Pole.
    The colored spheres represent modules which detected Cherenkov light from the charged particles of the event where the color signifies the relative time and the size of the sphere represents the relative light intensity seen by the module. 
    In this event, the earliest correlated hits (in yellow) occur on an inner string of the IceCube detector which is likely around the location of the neutrino interaction that produced a hadronic cascade and muon.
    The orange arrow shows the reconstructed path of the muon through the detector. 
    The light brown region indicates the dust layer--a region of the ice where scattering and absorption of photons is increased.}
    \label{fig:starting_event}
\end{figure}
ESTES was developed to select starting-track events created by muon neutrinos which undergo a charged-current (CC) weak-force interaction inside the IceCube detector geometry, producing a hadronic cascade in addition to the muon track.
A final level ESTES starting-track event is shown in Figure~\ref{fig:starting_event} to exemplify the morphological structure in data.
Identifying starting events allows for the distinction between astrophysical neutrinos and both atmospheric muons produced in cosmic-ray air showers, which deposit light as they enter into the detector, and atmospheric neutrinos that are accompanied into the detector by muons from the same air shower~\citep{Self_Veto:Schonert}.
Additionally, observing the interaction point of a muon neutrino also allows the energy reconstruction to estimate the muon energy at its conception and the energy deposited via inelastic scattering into the hadronic cascade, resulting in a more accurate energy reconstruction of the neutrino than for samples which use incoming muon events~\citep{IceCube:ESTESPRD}.
The final ESTES sample has a higher astrophysical neutrino proportion in the southern sky compared to other track selections, improving IceCube's sensitivity to some of the most active regions of the Galactic plane, such as the Galactic center, with track events.

The detailed motivation for a new starting-track event selection and a brief description of the selection and reconstruction techniques used by ESTES are provided in Section~\ref{sec:ESTES}.
Four neutrino source searches were performed using ESTES, each testing different hypotheses of neutrino production using a maximum likelihood approach.
The first test, detailed in Section~\ref{sec:AllSky}, is an all-sky scan that searched for statistically significant clusters of neutrinos using only the ESTES data by testing pixels across the whole sky.
The locations of Galactic and extra-galactic gamma-ray bright sources were tested individually, as reported in Section~\ref{sec:SrcList}.
Classes of Galactic sources--supernova remnants, pulsar wind nebulae, unidentified TeV objects, and TeV binaries--were also evaluated using a stacking analysis described in Section~\ref{sec:Stacking}.
Finally in Section~\ref{sec:GP}, an analysis was performed to look for diffuse neutrino emission from the Galactic plane created by the ambient cosmic-ray flux interacting with Galactic matter using the same template technique as~\citet{IceCube:DNNCSci}.
A discussion of the results is given in Section~\ref{sec:Discussion} and conclusions and future prospects of this work are presented in Section~\ref{sec:Conclusion}.

\section{Starting-Track Event Selection} \label{sec:ESTES}
\subsection{The starting-track event morphology in IceCube}
To be observed, neutrinos must interact and produce relativistic charged particles which deposit light via the Cherenkov effect.
The IceCube Neutrino Observatory uses a cubic kilometer of the naturally occurring glacial ice at the South Pole to observe the Cherenkov light from particles created in such neutrino interactions~\citep{IceCube:2017DetPaper}.
The main IceCube detector is comprised of digital optical modules (DOMs) which house photomultiplier tubes (PMTs) that are sensitive to single photons~\citep{IceCube:2017DetPaper}. 
The detector array consists of 86 strings, each spread approximately 125m apart in a hexagonal grid and instrumented with 60 DOMs from depths of 1450m to 2450m below the surface of the glacier~\citep{IceCube:2017DetPaper}. 
Additionally, an array of light-tight tanks, known as IceTop, sits on the surface of the glacier and can detect light from cosmic-ray air showers using PMTs frozen in ice~\citep{IceCube:2017DetPaper}.
The pattern of light deposition in the DOMs produced by the secondary charged particles, referred to as the event morphology, is used to infer information about the neutrino.
In IceCube, we can classify the morphologies generated by the charged particles into two general categories: track and cascade.
Tracks are produced by muons which come from cosmic-ray air showers, muon neutrino charged-current (CC) interactions, or tau neutrino CC interactions in which the tau decays to a muon with a 17.3\% branching fraction~\citep{PDG}.
A muon's path is relatively unaltered as it passes through the ice and, and muons with an initial energy above 1 TeV can traverse through the whole IceCube detector, leaving a clear track pointing back to their source and resulting in a resolution of the neutrino direction of around 0.5°.
The cascade morphology results from several types of neutrino interactions including neutral-current (NC) neutrino interactions, electron neutrino CC interactions, and tau neutrino CC interactions where the tau decays into an electron or hadrons. 
The particle showers produced from these interactions travel relatively short distances leading to a more isotropic distribution of light centered around the interaction vertex.
Cascade events are more poorly resolved than track events with resolutions of around 5°-15°~\citep{IceCube:DNNCSci}.
It should be noted that in the energy range which IceCube is sensitive to, neutrinos interact through deep inelastic scattering (DIS) meaning the neutrino will impart some energy to the nucleon producing a hadronic cascade including in the case of muon neutrino CC interactions~\citep{PDG}.
Tau neutrinos can produce other unique morphologies; however, those are not relevant in the energy range of focus in this paper ($\lesssim$ 100 TeV).

In this work, we focus on a sub-class of the track events, starting tracks (example event shown in Figure~\ref{fig:starting_event}), which are generated by muon (and tau) neutrino CC interactions where the interaction vertex is contained inside the detector geometry and light from a hadronic cascade and muon track are observed.
The other track events can be split into two other sub-classes: incoming tracks--where a muon is generated outside of IceCube's array, enters into the instrumented area, and most often leaves the detector--and skimming tracks--where a muon is generated outside of the detector and never enters the geometry but passes close enough to deposit light on the outer strings.
For incoming and skimming tracks, the energy reconstruction can only estimate the muon's energy at its first point of detection, effectively serving as a lower limit of the true neutrino energy.
The muon neutrino energy resolution is therefore improved when using starting tracks because the interaction vertex is observed, allowing the neutrino energy to be inferred from estimates of the muon's energy at production as well as the energy transferred to the hadronic component~\citep{IceCube:ESTESPRD}.

\begin{figure*}[tb!]
    \centering
    \includegraphics[width=0.49\textwidth]{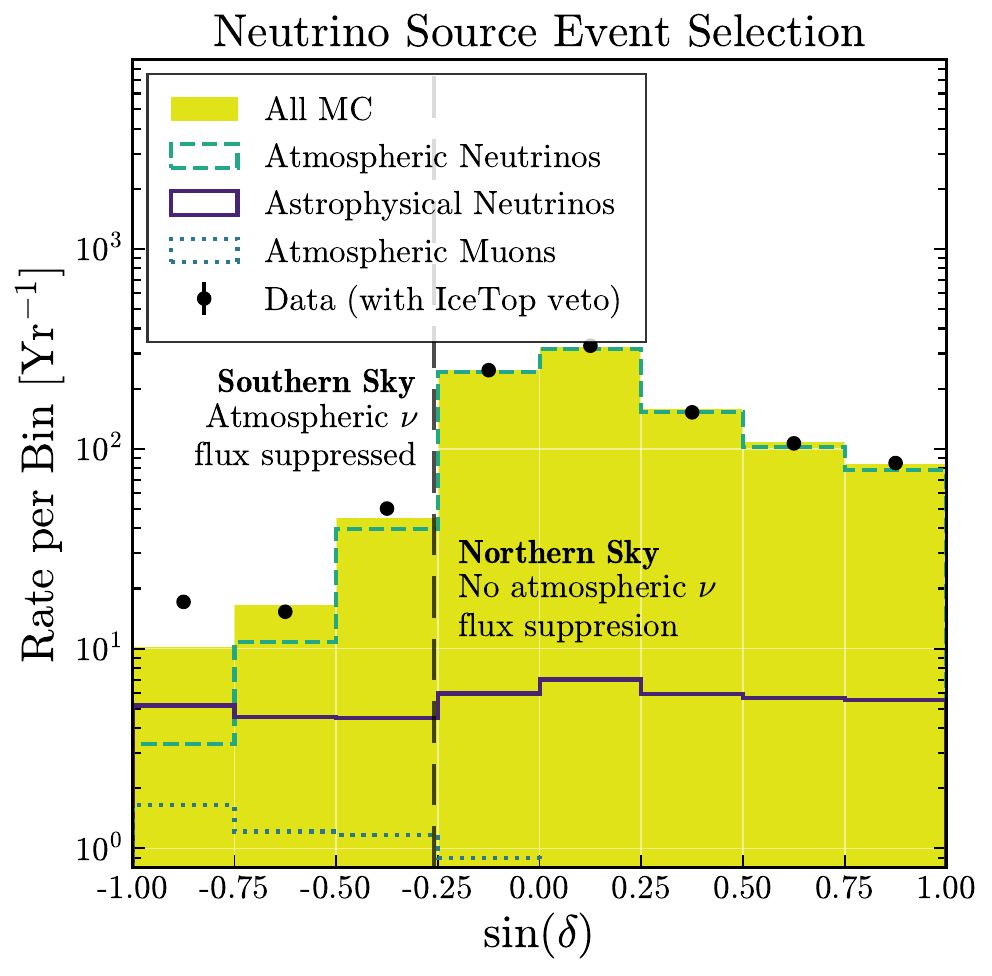}
    \hfill
    \includegraphics[width=0.49\textwidth]{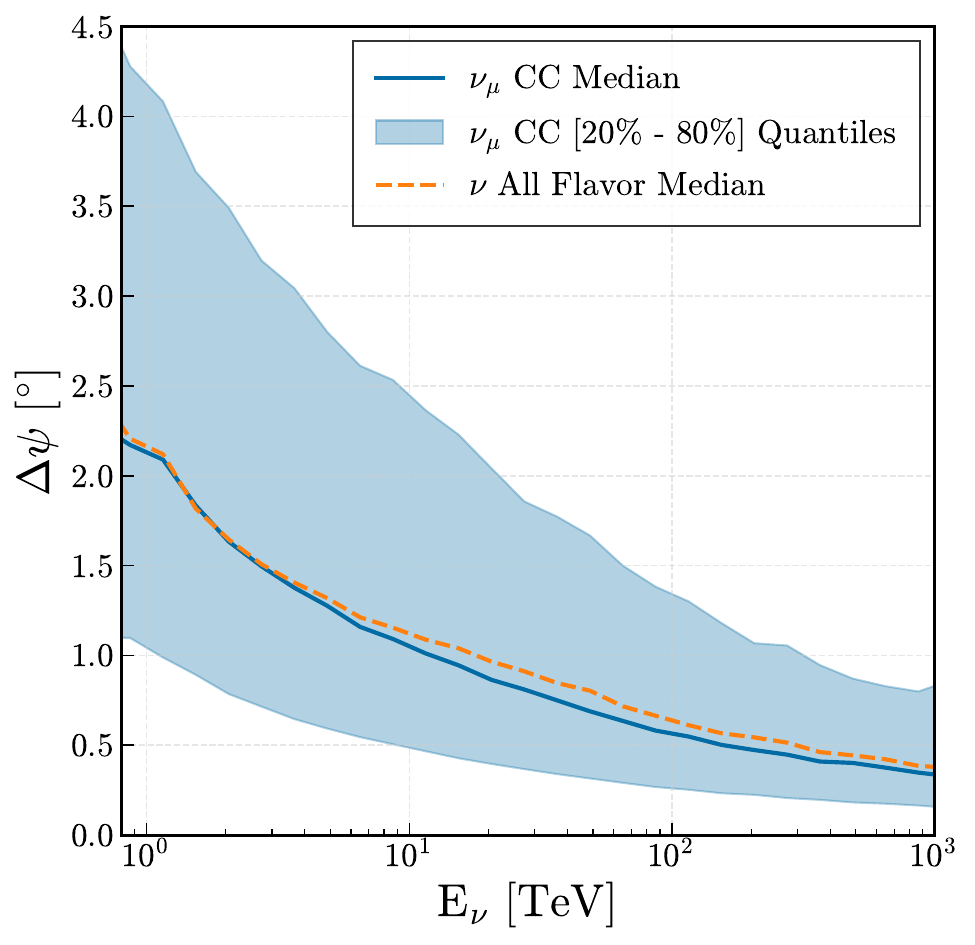}
    \caption{(Left) The final level event rates from data and simulation (MC) vs. the sine of the reconstructed declination of the events of the neutrino source searches version of ESTES. The MC is weighted according to the results of the ESTES diffuse astrophysical flux fit~\citep{IceCube:ESTESPRD}. The data is shown as black circles with small error bars. A disagreement between the data and MC was observed and is discussed in Subsection~\ref{Subsec:FinalES}. A suppression of the atmospheric-neutrino event rate is seen in the southern sky due to the atmospheric-neutrino veto that occurs when selecting starting events. Additionally, events which see in-time light in the IceTop array at the time the event would have passed through the surface detectors are removed from the data (for the simulation, joint IceCube-IceTop simulation was not used, so the cut could not be applied).
    (Right) The angular distance between the reconstructed and true direction, $\Delta\psi$, for MC events. The blue line represents the median angular distance as a function of true neutrino energy for muon neutrino charged-current interactions weighted by the astrophysical neutrino flux and the blue shaded regions contains the region of errors between the 20\% and 80\% weighted quantiles. The orange dashed lined shows the weighted-median angular error for all neutrino flavors and interaction types contained in the final level of the selection.}
    \label{fig:rates_errs}
\end{figure*}
The main motivating factor for a starting-track event selection is atmospheric muon and neutrino background rejection in the southern sky.
A simple approach to reject the 3 kHz muon background is to use the Earth as a veto and look only at events coming from the northern sky~\citep{IceCube:2021DiffuseNumu, IceCube:NGC1068Sci}.
However, with this method, the atmospheric-neutrino background still remains and is only distinguished from the astrophysical signal by the energy spectrum and zenith angle.
This method also removes the possibility of observing neutrinos from the southern sky where some interesting astrophysical objects, such as the Galactic center, lie.
The atmospheric muon rate in the southern sky can be reduced by adding charge cuts to select only the highest energy events, but the muon background contamination is still significant~\citep{IceCube:7yrPSTracks, IceCube:10yrPSTracks}.
Selecting for events that appear internally within the detector allows the rejection of atmospheric muons which produce light as they enter the instrumented volume.

The power of using starting events lies in the ability to not only reject atmospheric muons, but also atmospheric neutrinos which are often accompanied into the detector by muons from the same cosmic-ray air shower.
The probability of an atmospheric-neutrino event also including light from muons from the same air shower increases with the neutrino's energy, producing a noticeable effect in the rate of atmospheric neutrinos above 1 TeV~\citep{Self_Veto:Schonert}.
The atmospheric-neutrino veto effect can be seen in the left plot of Figure~\ref{fig:rates_errs}, where the atmospheric rate is suppressed for ``down-going'' events ($\sin (\delta) < -0.25$ where $\delta$ is the equatorial declination of the event).
This effect also pushes down the 1:1 atmospheric to astrophysical neutrino cross over-point in reconstructed neutrino energy from around 100 TeV to the tens of TeV depending on the zenith angle~\citep{IceCube:ESTESICRC2021}.
More detailed discussions of the modeling of the atmospheric-neutrino veto effect are given in~\cite{IceCube:ESTESPRD} which derived methods from \cite{Self_Veto:Gaisser, Self_Veto:Arguelles}.
Creating a specific starting-track sample has many uses, such as performing a fit of the diffuse astrophysical muon neutrino flux, which is described in~\cite{IceCube:ESTESPRD}.
Here we use the advantages of starting tracks in IceCube to investigate the southern sky for potential neutrino sources achieving an enhanced astrophysical neutrino purity from 1 - 500 TeV through the rejection of atmospheric muons and neutrinos.

\subsection{Design of the Enhanced Starting Track Event Selection (ESTES)}
ESTES looks to gather starting-tracks events, particularly in the southern sky with neutrino energies below 100 TeV, where many of IceCube's astrophysical muon-neutrino events have not yet been selected by other methods.
The previous starting event selections, such as the High-Energy Starting Events (HESE) sample~\citep{IceCube:HESE7yr}, are optimized to find high-energy cascades by using a fixed layer of in-ice DOMs as a veto, rejecting events where light is observed in the veto layer.  
In order to use the same fixed-veto-layer method, but lower the energy threshold, the fiducial volume must be substantially restricted as the atmospheric muon background is able to pass further through the detector without leaving a signal~\citep{IceCube:MESE}.

To identify starting-track events while maximizing the fiducial volume, a new tool was developed to reconstruct the position of the interaction vertex and calculate the probability that the event was an incoming or skimming muon, referred to as the starting-track veto (STV) ``miss probability''~\citep{IceCube:ESTESPRD, Mancina:2019ICRC}.
This tool uses parameterized simulations to estimate the light expected at IceCube DOMs based on a track hypothesis.
From the light estimates, the interaction vertex is reconstructed.
The ``miss probability'' quantifies the probability that the DOMs lying along the track before the vertex would have missed light from an incoming or skimming muon.
For example, an event which passes through an un-instrumented corridor of the detector will have a larger ``miss probability'' than one which passes close to the DOMs.
Similarly, for an event with a reconstructed vertex near the edge where it enters the detector, the ``miss probability'' will be much greater than for one where the vertex is located well within the detector.
The ``miss probability'' and other input variables are used to train a Boosted Decision Tree (BDT) to distinguish starting neutrino events from incoming or skimming atmospheric muons described in more detail in~\cite{IceCube:ESTESPRD}.

An observable unique to this work is the novel angular error estimator.
This parameter is used to model the signal Probability Distribution Function (PDF) in the likelihood test statistic.
Older IceCube analyses have estimated this parameter by using the shape of the directional reconstruction likelihood around the best fit point; however, for starting tracks we found the traditional estimators to be unreliable.
Therefore, we took an approach similar to the one used in~\cite{IceCube:NGC1068Sci} and used a Boosted Decision Tree (BDT) to create an angular error estimate for each event.
We used LightGBM~\citep{lightgbm} to train the BDT to reconstruct the angular distance between the true and reconstructed directions ($\Delta\psi$) with inputs such as the reconstructed zenith, the reconstructed length, the weighted location of light intensity registered by the DOMs of an event, and the reconstructed cascade and muon deposited energies.
The BDT output was then mapped to an angular error estimate by fitting the $\Delta\psi$ distribution with a von Mises function for events with a similar BDT output.
This estimator also worked well for distinguishing track-like events from cascade-like events.

For this work, a cut was applied to remove the cascade-like events by removing events with an angular error estimate greater than 10°.
A final cut was applied by removing events where the IceTop surface array measures a voltage in the PMTs greater than or equal to 2 photoelectron equivalents around the time the reconstructed track would have passed by the surface array.
This final IceTop cut adds an additional rejection for events likely to be atmospheric in origin and has been implemented in the other southern-sky track samples from IceCube~\citep{IceCube:3yrPSTracks2013,IceCube:10yrPSTracks}.

\subsection{Final ESTES event rate and comparison to other IceCube event selections}\label{Subsec:FinalES}
\begin{figure*}[tb!]
    \centering  
    \includegraphics[width=0.49\textwidth]{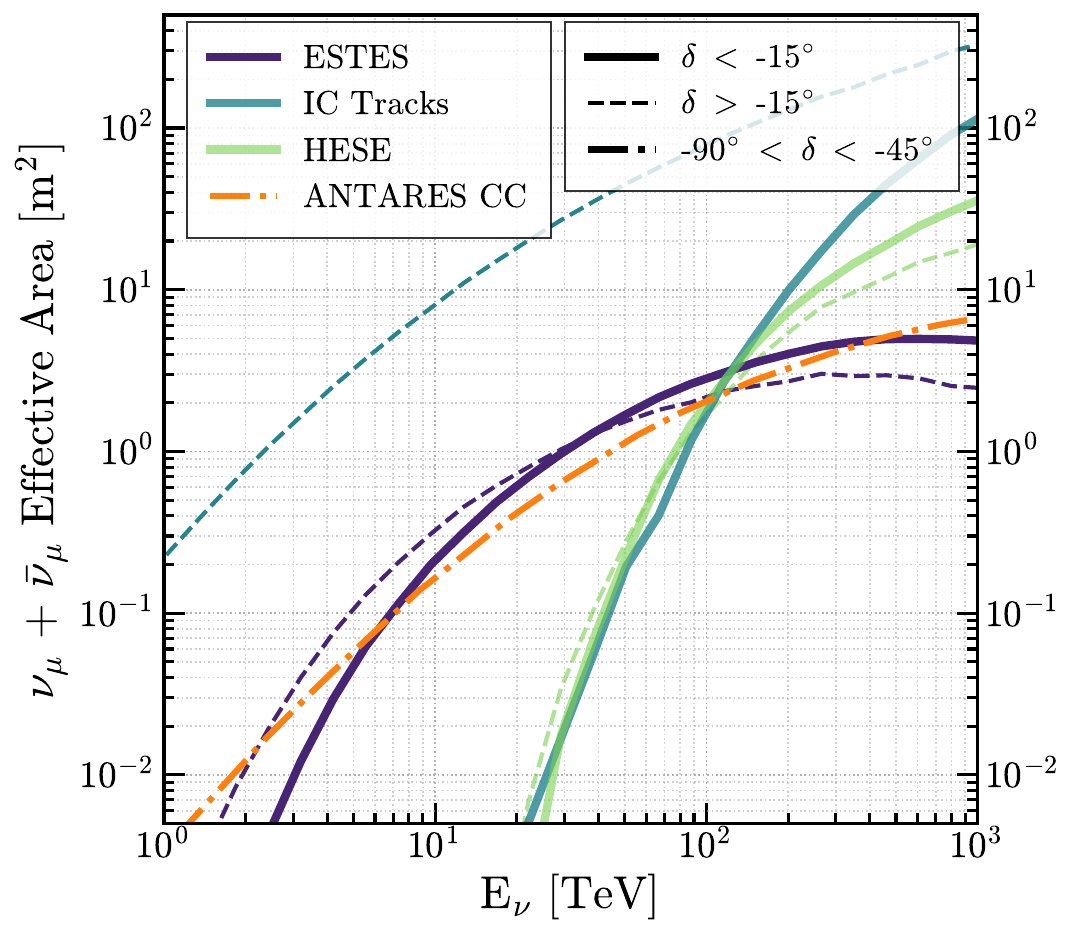}
    \hfill
    \includegraphics[width=0.49\textwidth]{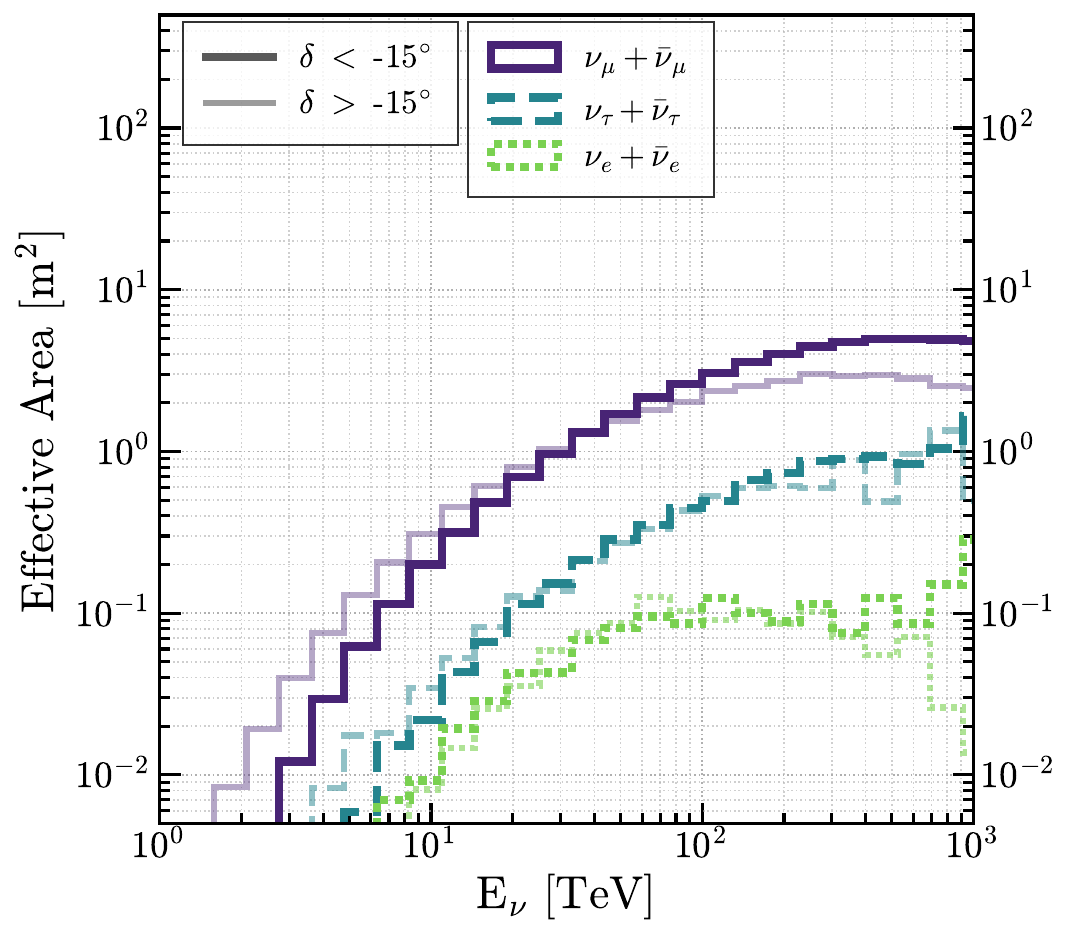}
    \caption{(Left) The muon neutrino and anti-neutrino effective area--a measurement of signal capture efficiency--versus the true neutrino energy for the ESTES selection (dark purple) compared to other selections including the IceCube tracks selection~\citep{IceCube:10yrPSTracks}, the High Energy Starting Events (HESE)~\citep{IceCube:HESE7yr}, and the ANTARES charged-current (CC) tracks~\citep{ANTARES:ANTARESFirstPS}. The different line styles represent different regions of the sky, motivating the focus of the ESTES analysis on the southern-equatorial sky. 
    (Right) The ESTES neutrino and anti-neutrino effective areas for the three neutrino flavors--distinguished by color and line-style--in two regions of the sky--denoted by line brightness and thickness.}
    \label{fig:effA}
\end{figure*}
The ESTES event selection was applied to 10.3 years of IceCube data and resulted in a sample of 10,350 total events, with 9,542 reconstructed as coming from the northern sky ($\delta \geq -15^{\circ}$) and 808 from the southern sky ($\delta < -15^{\circ}$)  where the atmospheric-neutrino veto effect occurs.
For most of this work, we differentiate the sky at $\delta$=15° as around this declination and for more negative declinations, the atmospheric-neutrino flux is suppressed by the veto effect.
From the simulation, around 63 astrophysical neutrinos, 520-530 low energy atmospheric neutrinos, and 30-50 atmospheric muons were expected in the southern sky assuming the best fit parameters from the diffuse neutrino flux fit by ESTES in~\cite{IceCube:ESTESPRD} (ranges given account for simulation statistics).
Figure~\ref{fig:rates_errs} shows the event rate, binned in sine declination, and the distribution of the angular distance between the true and reconstructed directions as a function of the true neutrino energy (right) for the final event selection.
In the rate plot (left of Figure~\ref{fig:rates_errs}), a disagreement between data and simulation rates is seen in the most southern bins.
Studies of the data excess in this region found events that appear by-eye to be clear starting tracks and not incoming or skimming muons. 
The discrepancy occurs only when the angular error estimate cut is applied which was not used in~\cite{IceCube:ESTESPRD}.
The cause of this disagreement when applying the angular error cut was traced to differences in the distribution of reconstructed directions for cascade-like events between simulation and data which may be due to incomplete modeling of the cascade events in simulation.
This data-simulation disagreement motivated the use of data-driven background estimation when creating the background probability distribution functions (PDFs) of the point-source likelihood and the background trials used in hypothesis testing.
The median angular distance between the true and reconstructed direction for all final level ESTES events is found to be 1.4°.
At the final level of the ESTES selection, the energy resolution was found to be within 25\%-30\% of the true neutrino energy in logarithmic space~\citep{IceCube:ESTESPRD}.

To compare ESTES to other samples of astrophysical neutrinos, we include their effective areas in the left plot of Figure~\ref{fig:effA}.
The IceCube tracks sample in~\cite{IceCube:10yrPSTracks} selects for all track-like events and in the southern sky requires events to be high energy ($\gtrsim$ 100 TeV) to reduce the atmospheric muon background rate.
The effect of this energy cut is the effective area for the tracks sample drops below the ESTES effective area around 100 TeV for the southern declinations.
It should also be noted that the atmospheric muon background still dominates the other tracks sample's astrophysical neutrinos by around 4 orders of magnitude in the southern sky at the final level selection compared to ESTES where the ratio is around 1 order of magnitude due to the atmospheric-neutrino veto~\citep{IceCube:10yrPSTracks, IceCube:7yrPSTracks}.

A search for overlapping events between ESTES and other IceCube samples was performed to understand how unique the events in the sample were.
In the northern sky (events with reconstructed declination, $\delta \geq -15^{\circ}$) a large overlap is found with the previous IceCube tracks samples~\citep{IceCube:10yrPSTracks, IceCube:NGC1068Sci}, which select for all track morphology sub-classes, as 71\% of the ESTES events are also captured in these samples.
On the contrary, in the southern sky ($\delta < -15^{\circ}$), only 1.8\% of ESTES events are found in the other tracks samples.
ESTES was found to have little overlap with the High Energy Starting Events (HESE) due to differences in the energy range and incoming-event veto method~\citep{IceCube:HESE7yr}, with 0.14\% of ESTES events found in the HESE sample across the whole sky.
Only 2.2\% of the ESTES events were in the DNN Cascades sample, responsible for the discovery of an excess of neutrinos from the galactic plane~\citep{IceCube:DNNCSci}.
From the overlap studies, it can be concluded that the ESTES sample provides a complementary and unique selection of relatively high-purity astrophysical muon-neutrino candidates from the southern sky.

\section{All-sky Search} \label{sec:AllSky}
\subsection{Method}
\begin{figure*}[tb!]
    \centering
    \includegraphics[width = 0.50\textwidth]{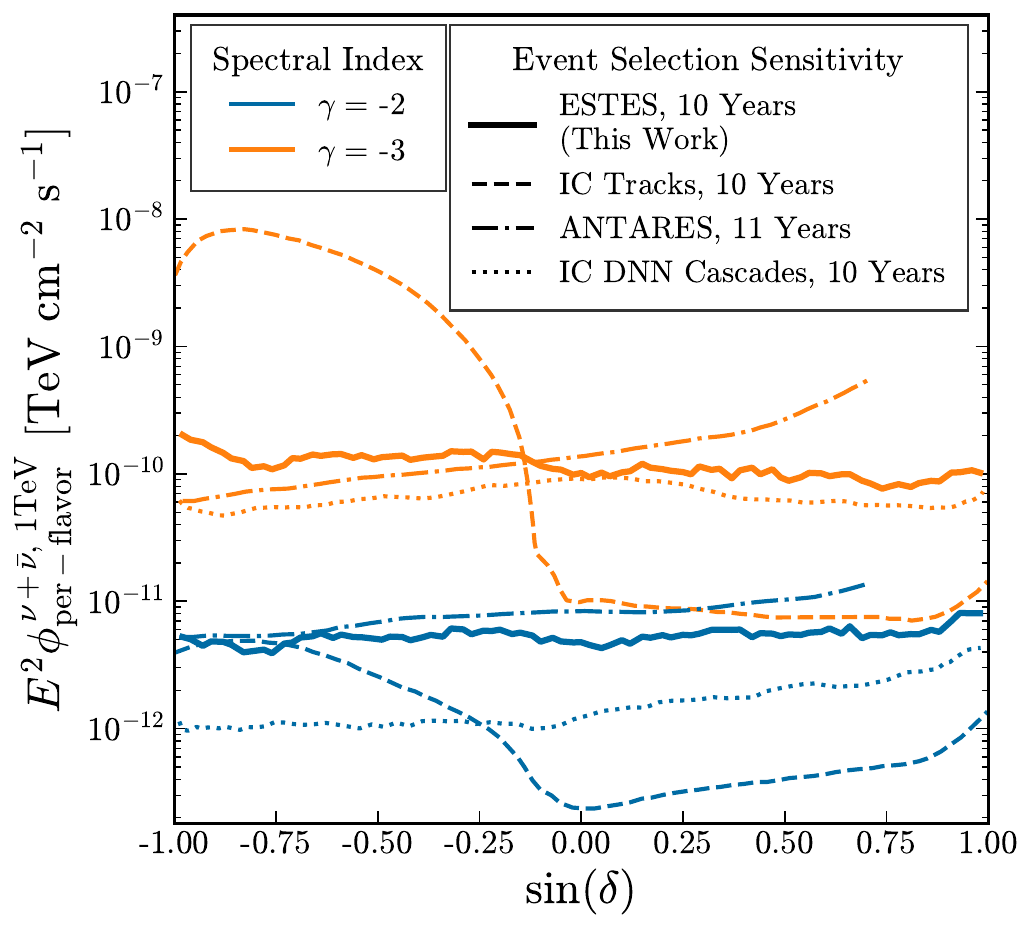}
    \hfill
    \includegraphics[width=0.48\textwidth]{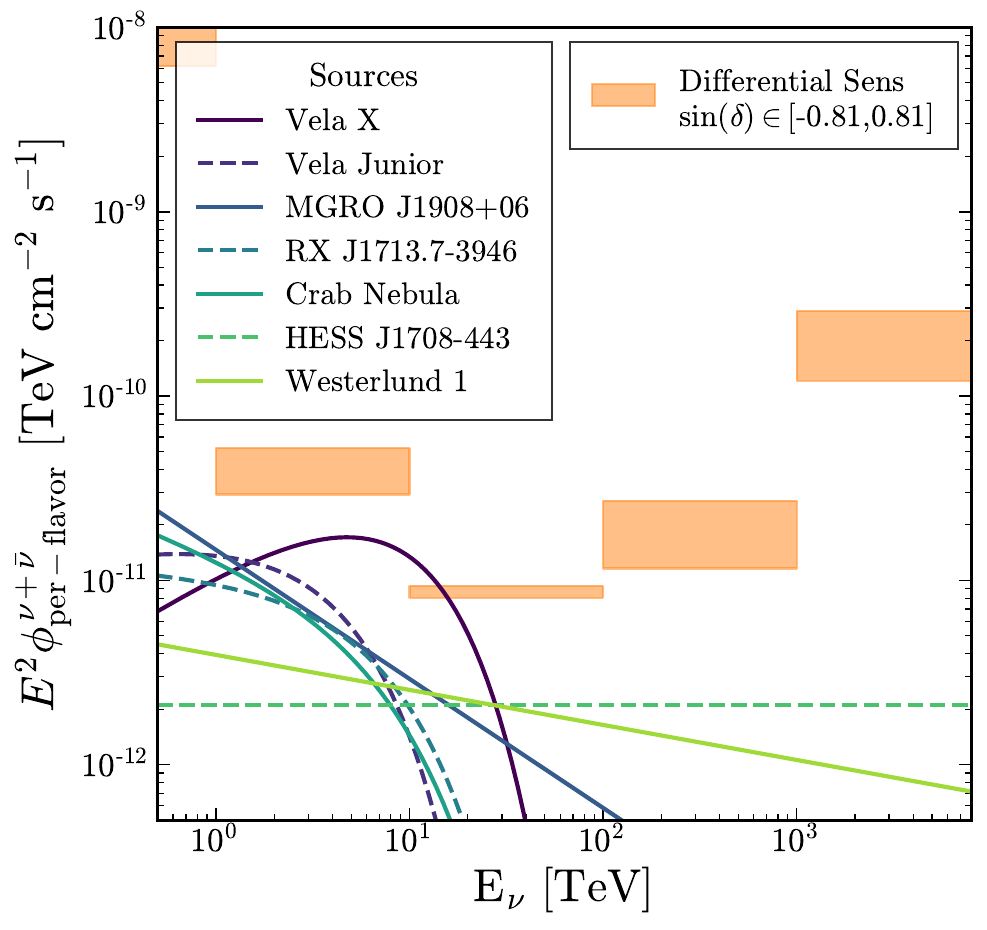}
    \caption{(Left) The sensitivity neutrino and anti-neutrino per-flavor flux at 1 TeV for a steadily emitting point source versus declination for the ESTES analysis (solid lines) for two different source spectral index hypotheses ($\gamma = -2.0$ in blue and $\gamma = -3.0$ in orange). The ESTES sensitivity is compared to that from the the 10 year IceCube tracks sample~\citep{IceCube:10yrPSTracks}, 11 year ANTARES~\citep{Aublin:Antares11Yr}, and the IceCube Deep Neural Network (DNN) selected Cascades search~\citep{IceCube:DNNCSci}.
    (Right) The maximum and minimum ESTES differential neutrino and anti-neutrino per-flavor sensitivity flux from simulations produced across a range of declinations (orange bands). The lines represent the models complied from sources in gamma-cat~\citep{GammaCat} and TeVCat~\citep{TeVCat} converted into a neutrino flux under an optimistic scenario~\citep{Halzen:2006AstroPartNu} for the 7 Galactic plane sources chosen in the source-list analysis.
    }
    \label{fig:sens}
\end{figure*}
The four analyses presented in this paper all use an unbinned maximum likelihood method~\citep{Braun:LLH} to generate test statistics to test against the same null hypothesis: that the ESTES events are produced by a flux isotropic in right ascension (accounting for the declination dependence of the atmospheric muon and neutrino backgrounds, the detector, and the ESTES selection). 
For point-source searches, the likelihood uses the angular separation between the sample's events and tested source location as well as the reconstructed energy of the events to estimate the number of signal neutrinos, $n_s$, and powerlaw spectral index, $\gamma$ of the hypothesized source, assuming a neutrino flux that follows a single power law: $\phi(E) = \phi_0 \left(\frac{E}{E_0}\right)^{\gamma}$.
The likelihood is formulated as:
\begin{equation}\label{eqn:PS_LLH}
    \mathcal{L}(n_s, \gamma; \vec{x}_s, \vec{x}_i, \sigma_i, E_i) = \prod_i^N \left( \frac{n_s}{N} \mathcal{S}(\vec{x}_s, \vec{x}_i, \sigma_i, E_i, \gamma) + (1 - \frac{n_s}{N}) \mathcal{B}(\delta_i, E_i) \right)
\end{equation}
where $\vec{x}_s$ is the hypothetical source location, $\vec{x}_i$ are the reconstructed best fit event locations comprised of declinations and right ascensions ($\delta_i$, $\alpha_i$), $\sigma_i$ are the reconstructed angular errors of each event, $E_i$ are the reconstructed energies of each event, and $N$ is the total number of events.
For the likelihood's signal probability distribution function (PDF), $\mathcal{S}$, we used a von Mises distribution to model the point spread function and used simulation to estimate the signal energy PDF as a function of declination.
The background spatial and energy PDFs, encompassed by $\mathcal{B}$ in equation~\ref{eqn:PS_LLH}, were generated by taking the spline of the data binned in reconstructed declination and energy.

A test statistic is generated using the log of the ratio of the likelihood maximized over $n_s$ and $\gamma$, $\mathcal{L}(\hat{n}_s, \hat{\gamma})$, to the likelihood assuming the null hypothesis, $\mathcal{L}(n_s=0)$.
Background simulations are produced by randomizing our data events in right ascension, under the assumption that most of our data events are from atmospheric muon and neutrino backgrounds.
The background test statistic distributions are calculated for several declinations ranging from -78.5° to 78.5°.
For the all-sky search, we test uniformly distributed pixels produced using healpy~\citep{Zonca2019, 2005ApJ:622:759G} across the sky, scanning for a point-like neutrino source. 
This provides an unbiased search for a point-like neutrino source; however, it is subject to significant trials corrections due to the nearly 800,000 pixels tested (healpy nside = 256 with a mean spacing of 0.23°).

To evaluate and compare the performance of ESTES for point-source searches, we produce sensitivities as a function of source location and energy range as shown in Figure~\ref{fig:sens}.
The sensitivity is defined as the median expected 90\% upper limit flux found in the absence of an observed signal from simulations.
In the left plot of Figure~\ref{fig:sens}, the sum of the neutrino and anti-neutrino per-flavor sensitivity flux at 1 TeV is given for ESTES and other neutrino samples that performed sources in the southern sky.
We find that ESTES improves on the tracks sample~\citep{IceCube:10yrPSTracks} in the southern sky for sources with a spectral index more negative than 2, and it is comparable to the ANTARES 11 years sensitivity~\citep{Aublin:Antares11Yr} and the IceCube cascades~\citep{IceCube:DNNCSci} samples.
The ESTES sample also improves on the previous sensitivity using IceCube tracks for sources with exponential energy cutoffs in the southern sky, which is expected from the effective area comparison in Figure~\ref{fig:effA}.

In the right plot of Figure~\ref{fig:sens}, the differential sensitivity flux for different decades in energy is given assuming a spectral index of $\gamma$ = 2.0 in limited energy range.
The differential sensitivity is represented by an orange band which give the minimum and maximum flux values calculated over a range of declinations for each decade across the whole sky.
The ESTES sample is most sensitive between 10 and 100 TeV.
The sensitivity is compared to models of the neutrino flux from TeV gamma-ray bright Galactic plane sources which will be discussed further in Section~\ref{sec:SrcList}.

\subsection{Results}
\begin{figure*}[tb!]
    \centering
    \includegraphics[width=0.8\textwidth]{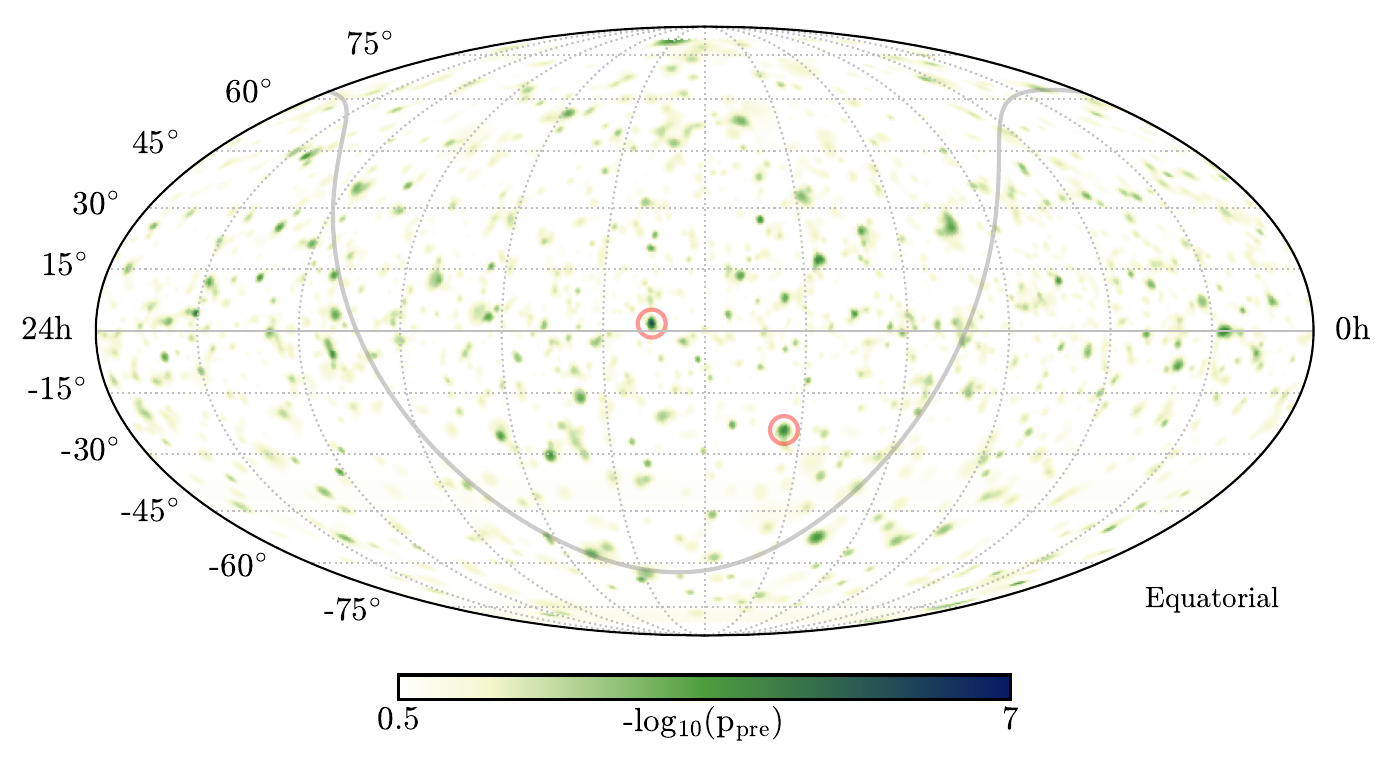}
    \caption{Map of the ESTES all-sky scan pre-trial p-values, p$_{\mathrm{pre}}$. The hottest spots found in the northern and southern equatorial sky are highlighted by the red circles.}
    \label{fig:allskyresult}
\end{figure*}
The resulting pre-trial p-values for each pixel are shown in Figure~\ref{fig:allskyresult}.
We highlight the brightest points found in the northern and southern sky at $(\delta, \alpha) = (+1.79, 195.64)$ and $(\delta, \alpha) = (-23.97, 155.21)$ respectively.
The northern hotspot results in the most significant pre-trial p-value of $1.4\times10^{-6}$.
However, after correcting for the number of pixels tested by performing full scans of background simulation maps to account for the angular error radius being larger than the map resolution, the post-trial p-value is found to be 0.22.
Therefore, the null hypothesis cannot be rejected in the all-sky scan analysis.

\section{Search for Neutrinos from Selected Gamma-Ray Sources} \label{sec:SrcList}
\subsection{Method}
The second analysis performed looked for significant clusters of neutrinos from the direction of known gamma-ray sources.
The source selection was based on the hypothesis that astrophysical neutrinos and gamma-rays are both produced via pion decay from high energy cosmic rays interacting with the matter or radiation in the vicinity of the cosmic-ray accelerator.
Under this assumption, we selected sources that are bright in gamma-rays taking into account the declination dependence of the ESTES sample, which was the same approach detailed in~\cite{IceCube:10yrPSTracks}.
To select extra-galactic sources, we used the flux at 1 GeV for sources reported in the $Fermi$-LAT 4FGL catalog to select the 103 brightest sources~\citep{Fermi-LAT:Catalog}.
For Galactic sources, the TeV gamma-ray data compiled in TeVCat~\citep{Wakely:TeVCat,TeVCat} and GammaCat~\citep{GammaCat} was converted to a neutrino flux assuming the optimistic scenario that all gamma-ray emission was produced by proton-proton collisions at the source~\citep{Halzen:2006AstroPartNu}.
The Galactic sources were then compared to the ESTES differential sensitivity at the sources' declination, illustrated in the right plot of Figure~\ref{fig:sens}, and the top 7 sources were selected.
The final number of 110 Galactic and extra-galactic sources was chosen such that a 5$\sigma$ pre-trial significance would drop to a 4$\sigma$ post-trial significance.
All sources were treated as point-like neutrino emitters in the likelihood evaluation.

\subsection{Results}
The most significant source identified by the ESTES sample was 1ES 0647+250, a BL Lac located in the northern sky.
The pre-trials correction p-value is $4.1 \times 10^{-4}$ and, after correcting for the 110 source locations tested, the post-trial p-value was found to be 0.044.
Because this source is located in the northern sky, the previous track analyses have a greater sensitivity to 1ES 0647+250 and also tested the location.
In the previous analysis with IceCube tracks, no signal neutrinos were found from the direction of the source and a p-value of 0.53 was reported~\citep{IceCube:10yrPSTracks}.
The previous analysis sets a more constraining 90\% upper limit of $3.0\times10^{-13}$ TeV$^{-1}$ cm$^{-2}$ s$^{-1}$ assuming a powerlaw energy spectrum with $\gamma = -2$~\citep{IceCube:10yrPSTracks} (for this work the upper limit calculated was $1.7\times10^{-11}$ TeV$^{-1}$ cm$^{-2}$ s$^{-1}$).

\begin{figure*}[tb!]    \centering
    \includegraphics[width=0.5\textwidth]{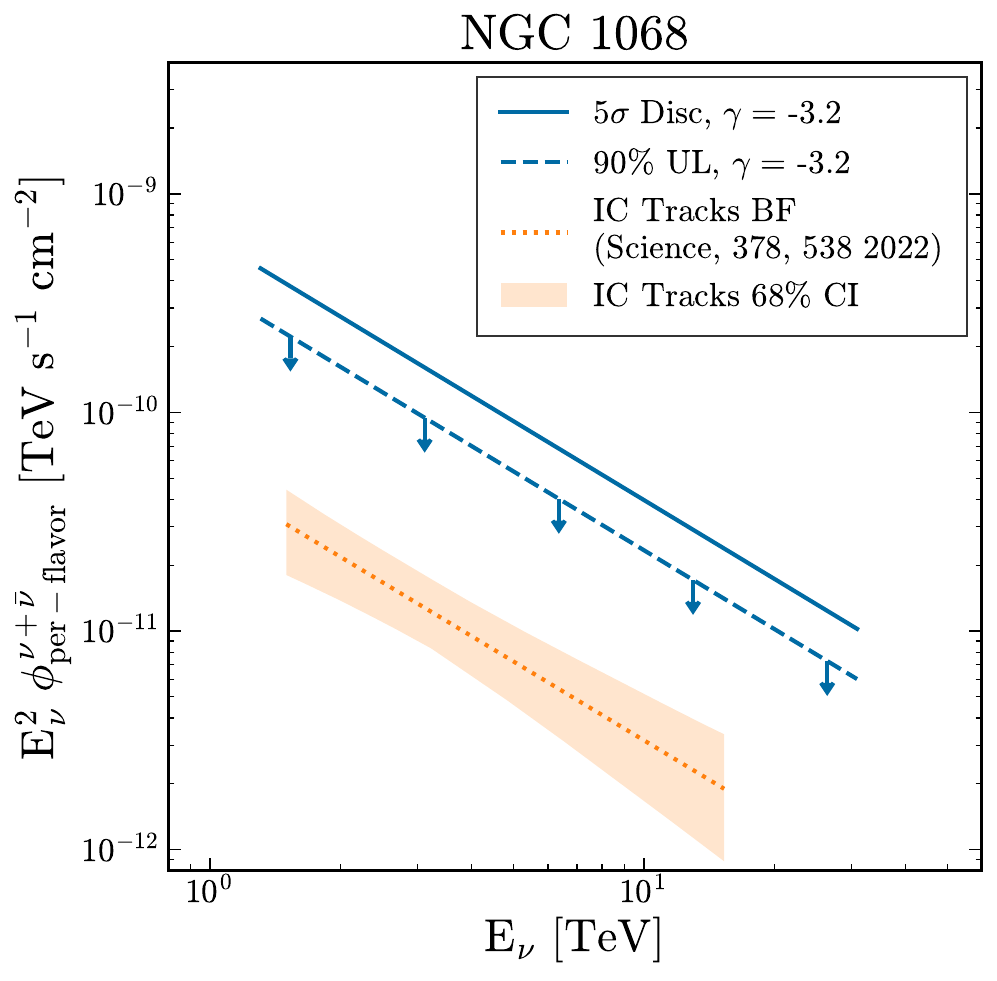}
    \caption{The 5$\sigma$ discovery potential (Disc) and 90\% upper limit (UL) neutrino and anti-neutrino per-flavor flux from ESTES, this work, (blue) compared to the best-fit (BF) and confidence interval (CI) fluxes from the NGC 1068 discovery from~\cite{IceCube:NGC1068Sci} (orange). The energy range shown for this work was selected by finding the central range in which 90\% of simulated source events lie assuming the tracks best fit powerlaw spectrum of $\gamma = -3.2$~\citep{IceCube:NGC1068Sci}. The tracks energy range was selected by defining the central range where 68\% of the top contributing data events lie.}
    \label{fig:ngc1068_senslims}
\end{figure*}

The full results for all sources are given in Table~\ref{tab:SSCatalogResults} in Appendix~\ref{App:SrcListTables}.
The most significant source in the southern region of the sky ($\delta < -15^{\circ}$), where ESTES provides the most unique events, was the Small Magellanic Cloud (SMC) which reports a pre-trial p-value of 0.03.
NGC 1068 and TXS 0506+056, two sources with evidence of neutrino emission~\citep{IceCube:NGC1068Sci, IceCube:TXSMMPaper, IceCube:TXSNeutFlare}, are the 10th and 15th most significant sources in our catalog.
Both sources lie above a declination of -15°, where the previous track analyses are more sensitive than ESTES (illustrated in the right plot of Figure~\ref{fig:sens}). 
In Figure~\ref{fig:ngc1068_senslims}, the ESTES 5$\sigma$ discovery potential and 90\% upper limits assuming the best fit spectral index for NGC 1068 from~\cite{IceCube:NGC1068Sci} analysis are shown to both lie above the 68\% confidence interval fit by the tracks sample.
The 5$\sigma$ discovery potential is defined as the flux where 50\% of injection trials are greater than the 5$\sigma$ test statistic from the background trials distribution.
Additionally, many northern sky ($\geq$ 15°) ESTES events are also captured in the other track samples; in fact, of the 8 most significant events contributing to the ESTES NGC 1068 fit, 6 are also found in the sample from~\cite{IceCube:NGC1068Sci}. 
It is concluded that no new neutrino sources were identified by this work.

\section{Galactic Plane Source Stacking Catalogs} \label{sec:Stacking}
\subsection{Method}
To test the correlation of the ESTES events with specific classes of known sources, a stacking analysis was performed.
To ``stack'' a catalog, the log-likelihood values at the locations of similarly classified sources are summed to produce a single test statistic.
For ESTES, Galactic plane sources were of particular focus because many lie in the southern sky and have been observed to have energy spectra with powerlaw indices less than -2.0 and energy cutoffs in the single to tens of TeV when measured in TeV gamma-rays~\citep{GammaCat, TeVCat}.
Sources from the TeVCat and GammaCat databases~\citep{Wakely:TeVCat, TeVCat, GammaCat} were placed into four catalogs: supernova remnants (SNR), pulsar wind nebulae (PWN), TeV binaries (BIN), and unidentified TeV gamma-ray sources (UID).
The UID sources are bright sources in TeV gamma-rays for which the associated source at other wavelengths has not been concretely identified, but many of the sources are likely to be either SNR or PWN.
For the SNR, PWN, and UID catalogs, the top 12 brightest gamma-ray sources of each type were included; whereas, the BIN catalog included all 7 TeV binaries that had been identified at the time.

Though the sources were selected based on gamma-ray brightness, the log-likelihood values evaluated at each source location were summed together without applying any weights to favor the brighter sources and we fit a single floating spectral index, $\gamma$, for each catalog.
However, in deriving a sensitivity, discovery potential, and upper limits a model for the neutrino emission was used for the PWN, SNR, and UID catalogs.
The models for each source were derived using the same optimistic approach as outlined for the Galactic sources in Section~\ref{sec:SrcList}.
The parameters for the neutrino flux models from gamma-ray data assuming a proton-proton model for the PWN, SNR, and UID sources are given in Table~\ref{tab:GPStackingResults_SNRPWNUID} in Appendix~\ref{App:SrcListTables} and are shown as the faint orange dotted lines in Figure~\ref{fig:stackuplims}.
The ESTES analysis was found to be slightly sensitive to the PWN and UID catalogs according to these models, but not sensitive to the SNR catalog under the model hypothesis.
No models were derived for the BIN catalog because of the time-variability of the objects' flux.
Note that, due to the location of the sources of these catalogs, the signal hypothesis tested is correlated with the diffuse Galactic plane neutrino hypothesis test described in Section~\ref{sec:GP}.

\subsection{Results}
\begin{figure*}[tb!]
    \centering
    \includegraphics[width=\textwidth]{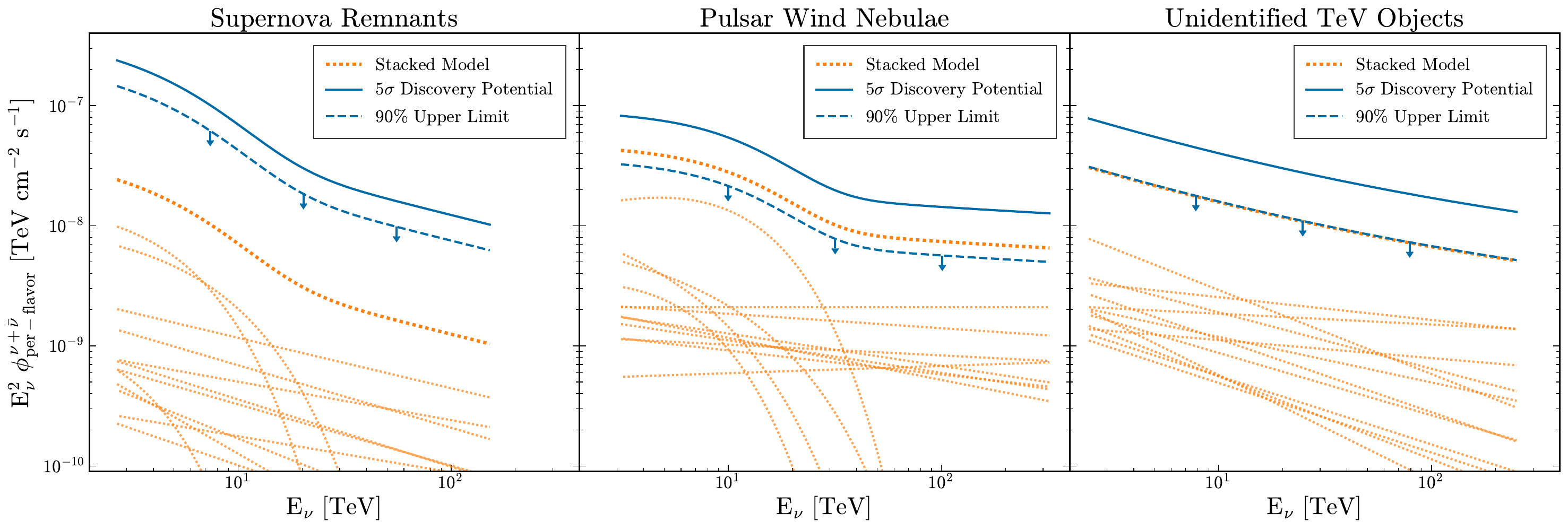}
    \caption{The 5$\sigma$ discovery potential and 90\% statistics-only upper limit neutrino and anti-neutrino per-flavor fluxes for the stacking catalogs for the three catalogs compared to the model neutrino fluxes (blue lines). The source models (orange, dotted lines) were generated by compiling the TeV gamma-ray data from GammaCat~\citep{GammaCat} and TeVCat~\citep{TeVCat} and converting the data into a neutrino flux assuming an optimistic scenario where all TeV gamma-rays are produced from pions originating from proton-proton interactions~\citep{Halzen:2006AstroPartNu}. The 90\% upper limits set for the pulsar wind nebulae rule out 100\% of the model flux. The energy range reported represents the central range where 90\% of simulated source neutrinos lie assuming the model flux hypothesis.}
    \label{fig:stackuplims}
\end{figure*}
The most significant catalog evaluated in the stacking analysis was the SNR catalog with a pre-trial p-value of 0.040. 
To correct for trials, we collected the best fit TS from background simulations where the four stacking catalog analyses were run on the same map to account for correlations between the fits to each catalog.
The post-trial p-value was found to be 0.056 using this technique; therefore, the null hypothesis could not be rejected.
The results and upper limits for all four catalogs are given in Tables~\ref{tab:GPStackingResults_SNRPWNUID} and~\ref{tab:GPStackingResults_BIN} of Appendix~\ref{App:SrcListTables}.
Using the gamma-ray derived models, the 90\% upper limits were calculated for the SNR, PWN, and UID catlogs and are compared to the model expectation fluxes in Figure~\ref{fig:stackuplims}.
For the PWN and UID catalog the 90\% upper limits are 77\% and 101\% of the model flux.
For the SNR catalog, the upper limit lies at 601\% of the model flux and is much less constraining due to the weaker predicted flux from the models and the fitted result.
The BIN catalog returned a TS of 0.0, fitting no signal neutrinos from the direction of the TeV binary sources, and therefore, competitive upper limits can be set on emission from TeV Binaries.
Assuming equivalent fluxes from each source and a powerlaw spectral energy distribution of $\gamma$ = -2.0 and -3.0, the 90\% statistics-only upper limits are respectively $3.0\times10^{-12}$ TeV$^{-2}$ cm$^{-2}$ s$^{-1}$ and  $9.2\times10^{-13}$ TeV$^{-2}$ cm$^{-2}$ s$^{-1}$.

\section{Diffuse Galactic Plane Neutrino Emission} \label{sec:GP}
\subsection{Method}
A flux of cosmic rays permeates our Galaxy and interacts with the interstellar medium, producing a diffuse flux of neutrinos and gamma-rays from the Galactic plane~\citep{Gaggero:KRAgModelPaper, IceCube:DNNCSci}.
Two models of diffuse Galactic plane neutrinos are tested here: a model extrapolated from the $Fermi$-LAT $\pi^0$ model for diffuse galactic gamma rays~\citep{Fermi:FermiPi0Model} and the KRA$_{\gamma}$ model~\citep{Gaggero:KRAgModel, Gaggero:KRAgModelPaper}.
Both models were produced by injecting a cosmic-ray flux into propagation tools which interact it with the interstellar medium (ISM) which is modeled by hydrogen I and CO emission line observations~\citep{Fermi:FermiPi0Model}.
The two models' parameters are fitted using the locally observed cosmic-ray data and are validated by gamma-ray observations~\citep{Fermi:FermiPi0Model, Gaggero:KRAgModelPaper}.
The KRA$_{\gamma}$ model includes a Galactic radial dependence in the cosmic-ray diffusion coefficient, resulting in an emission intensity map more peaked around the Galactic center than the $Fermi$-LAT $\pi^0$ model~\citep{Gaggero:KRAgModel, Gaggero:KRAgModelPaper}.
Additionally, a per nucleon exponential energy cutoff is applied to the cosmic-ray spectra in the KRA$_{\gamma}$ model and the authors produce a prediction for the neutrino spectrum~\citep{Gaggero:KRAgModel, Gaggero:KRAgModelPaper}. 
Besides the shape differences in the energy spectrum and relative morphological shape of the models, the KRA$_{\gamma}$ model also predicts a larger expected diffuse Galactic plane neutrino flux at TeV energies than the extrapolation of the $Fermi$-LAT $\pi^0$ model (which is observed by comparing the dotted lines in the plots of Figure~\ref{fig:GP_upperlimits})~\citep{Gaggero:KRAgModel, Gaggero:KRAgModelPaper}.

To test these models, the maximum likelihood approach was again used, but to account for the large size of the emission region a ``signal-subtraction'' component was added to remove signal contamination from the background PDFs generated from data.
This same likelihood technique and galactic plane models were used in previous searches for Galactic diffuse neutrino emission using different IceCube event selections~\citep{IceCube:PSTracks7YRGP, IceCube:DNNCSci}.
Unlike the point-source tests described above, the energy spectrum was fixed.
In the analysis of the $Fermi$-LAT $\pi^0$ model, a single powerlaw distribution with a spectral index of $\gamma$=-2.7 was used to describe the energy spectrum.
Meanwhile, the likelihood of the KRA$_{\gamma}$ model incorporated the model's predicted spectra, testing two exponential energy cutoffs applied to the cosmic-ray flux at 5 PeV and at 50 PeV~\citep{Gaggero:KRAgModel}.
The sensitivity flux for these models are shown for ESTES and compared to previously published results using the same analysis method in Table~\ref{tab:GPResultsCompare}.
ESTES is more sensitive than all other previous analyses using muon track events in the southern sky in IceCube and ANTARES~\citep{IceCube:PSTracks7YRGP, ANTARES:IceCube:GP, ANTARES:KRAgTemplate2023}, but it is less sensitive than the DNN cascades sample which found an 4.5$\sigma$ excess of neutrinos correlated with the $Fermi$-LAT $\pi^0$ model~\citep{IceCube:DNNCSci}.

\subsection{Results}
\begin{figure*}[tb!]
    \centering
    \includegraphics[width=\textwidth]{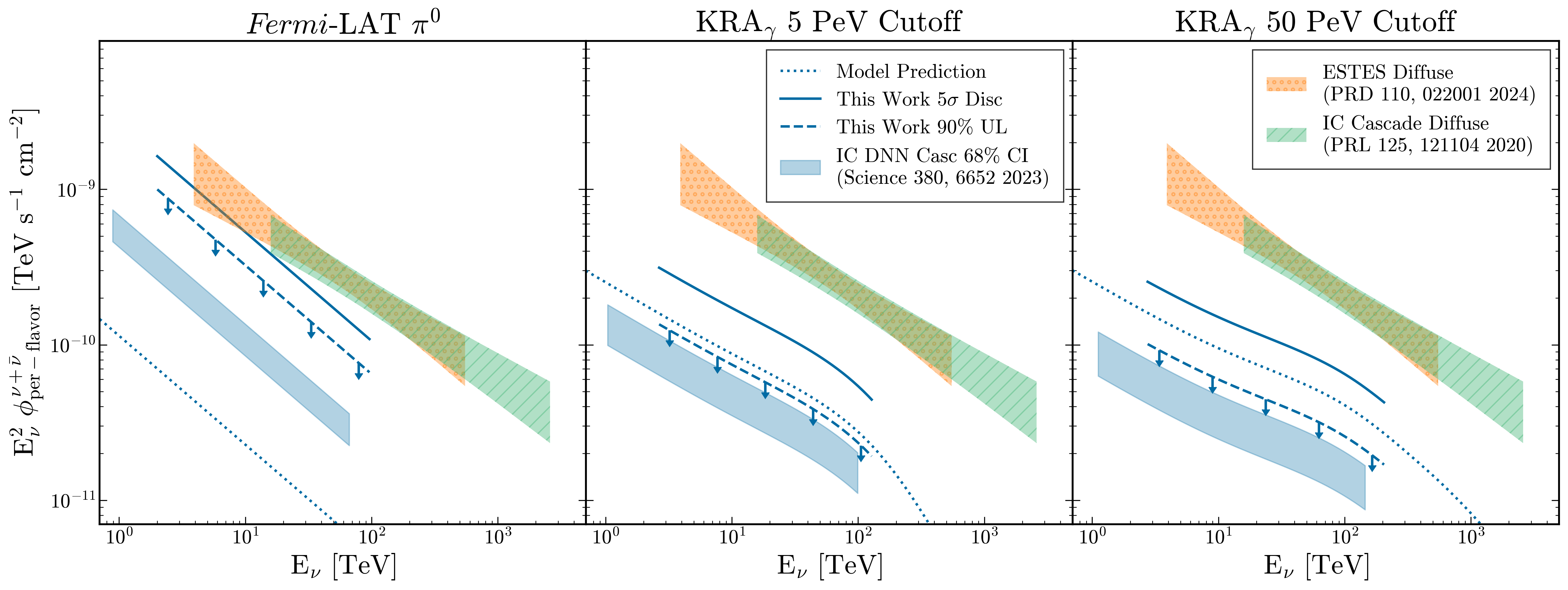}
    \caption{The 5$\sigma$ discovery potential (Disc) and 90\% statistics-only upper limit (UL) neutrino and anti-neutrino per-flavor flux for the ESTES analyses compared to the confidence intervals (CI) from the IceCube DNN Cascades (IC DNN Casc) analysis~\citep{IceCube:DNNCSci} for the three Galactic plane template models: $Fermi$-LAT $\pi^0$ model~\citep{Fermi:FermiPi0Model}(left) and the KRA$_{\gamma}$~\citep{Gaggero:KRAgModel} with a 5 PeV (middle) and 50 PeV (right) cosmic-ray exponential cutoff. The ESTES UL rule out the KRA$_{\gamma}$ model prediction at 90\% confidence~\citep{Gaggero:KRAgModel}. For added comparison, the astrophysical diffuse fits for ESTES ~\citep{IceCube:ESTESPRD} and the diffuse cascade selection~\citep{IceCube:SCDiffuse} are included. The energy range shown for this work is defined by simulating the source hypotheses and finding the central range where 90\% of the simulated events lie. The energy range for the DNN Cascades is calculated from the central range where 90\% of the contributing data events lie.}
    \label{fig:GP_upperlimits}
\end{figure*}

\begin{table*}[tbh!]
    \centering
    \begin{tabular}{|r||c|c|c||c|c|c||c|c|c|}
         \hline
          \multicolumn{1}{|r||}{} & \multicolumn{3}{c||}{$Fermi$-LAT $\pi^0$, $\gamma$ = -2.7} & \multicolumn{3}{c||}{KRA$_{\gamma}$ 5 PeV Cutoff} & \multicolumn{3}{c|}{KRA$_{\gamma}$ 50 PeV Cutoff} \\
          \hline
         Event Selection & $\phi_{90\%}^{\mathrm{sens}}$ & $\phi_{90\%}^{\mathrm{UL}}$ &  $-\log_{10}(p)$ & MF$_{90\%}^{\mathrm{sens}}$ & MF$_{90\%}^{\mathrm{UL}}$ &  $-\log_{10}(p)$ & MF$_{90\%}^{\mathrm{sens}}$ & MF$_{90\%}^{\mathrm{UL}}$ & $-\log_{10}(p)$\\
         \hline
         ESTES (This Work) & 2.6 & 6.5 & 1.4 & 53\% & 85\% & 0.78 & 47\% & 63\% & 0.75 \\
         \hline
         DNN Cascades & 0.598 & -- & 5.9 & 16\% & -- & 5.2 & 11\% & -- & 4.4 \\
         \hline
         IC Tracks, 7 Years & -- & -- & -- & -- & -- & -- & 79\% & 120\% & 0.54 \\
         \hline
         ANTARES and IC Tracks & -- & -- & -- & 81\% & 119\% & 0.54 & 57\% & 90\% & 0.58 \\
         \hline
         ANTARES 2023 & -- & -- & -- & 93\% & 199\% & 1.31 & -- & -- & -- \\
         \hline
    \end{tabular}
    \caption{Comparison of the ESTES Galactic Plane results to the IceCube Deep Neural Network (DNN) selected Cascades~\citep{IceCube:DNNCSci}, 7 years of IceCube tracks~\citep{IceCube:PSTracks7YRGP}, the joint ANTARES and IceCube tracks analysis~\citep{ANTARES:IceCube:GP}, and the ANTARES 2023 results~\citep{ANTARES:KRAgTemplate2023}. The three models tested were the $Fermi$-LAT $\pi^0$ model~\citep{Fermi:FermiPi0Model} assuming a powerlaw spectral index of $\gamma=-2.7$ and the KRA$_{\gamma}$ model~\citep{Gaggero:KRAgModel} assuming a 5 PeV and 50 PeV exponential energy cutoff on the cosmic-ray spectrum.
    For the $Fermi$-LAT $\pi^0$ model, the 90\% sensitivity and statistics-only upper limits, $\phi_{90\%}^{\mathrm{sens}}$ and $\phi_{90\%}^{\mathrm{UL}}$, are given as E$^2$ $\phi^{\nu+\bar{\nu}}_{\mathrm{per-flavor}}$ at 100 TeV reported in units of $\times 10^{-11}$ TeV cm$^{-2}$ s$^{-1}$.
    For the KRA$_{\gamma}$ models, the sensitivity and upper limits are given as a percentage of the model flux (MF) given by the reference~\citep{Gaggero:KRAgModel}. The p-values reported here are the pre-trial's corrected p-values.
    }
    \label{tab:GPResultsCompare}
\end{table*}
The analysis returned a most significant pre-trial p-value of 0.040 for the $Fermi$-LAT $\pi^0$ model, and, after correcting for three maps, found a post-trial p-value of 0.057.
Therefore the null hypothesis cannot be rejected by the ESTES result alone.
The preference for the $Fermi$-LAT $\pi^0$ model appears to be motivated by the spatial distribution of events as no neutrinos with energies greater than 10 TeV are found in the direction of the Galactic center in the ESTES sample.
However, this preference is not statistically significant.
The 90\% upper limits for ESTES are given in Table~\ref{tab:GPResultsCompare} and shown in Figure~\ref{fig:GP_upperlimits}.
The KRA$_{\gamma}$ 90\% statistics-only upper limits rule out 100\% of the model flux for both choice of cutoff; however, they lie above the 68\% confidence intervals fit by the DNN cascades analysis and the upper limit calculations do not include modeling of systematic uncertainties~\citep{IceCube:DNNCSci}.

\section{Discussion}\label{sec:Discussion}
We performed four analyses to search for sources of astrophysical neutrinos.
The all-sky scan, source-list search, Galactic plane source stacking catalogs, and diffuse Galactic plane templates all returned significances of less than 2$\sigma$ when accounting for the trials performed in each of the four analysis, so no new hints of the origin of neutrino sources were identified.
A global trials corrected p-value was not calculated accounting for the four types of analyses because none of the values were statistically significant after accounting for the trials within each analysis type.
However, ESTES was able to provide the most constraining upper limits from IceCube for many Galactic sources of neutrinos.

In 2023, IceCube announced the discovery of Galactic plane neutrinos with the DNN cascades sample with 4.5$\sigma$  post-trial significance~\citep{IceCube:DNNCSci}.
The DNN cascade result had a preference for the $Fermi$-LAT $\pi^0$ template as a model for the emission; however, the other models of Galactic plane emission tested--the KRA$_{\gamma}$ template and stacking catalogs--all rejected the null hypothesis with a significance greater than 3$\sigma$.
Discerning between different models of Galactic plane neutrino emission is difficult in this early era of neutrino astronomy, especially for cascade events with a median angular resolution between 5°-15°.
The ESTES sample has an improved resolution of around 1.4° but a smaller effective area, meaning that, on its own, this sample could not reject the null hypothesis.
The 90\% statistics-only upper limits presented here limit the KRA$_{\gamma}$ to be less than 100\% of the predicted model flux for both cutoff energy hypotheses.
The ESTES results also rule out an optimistic scenario for neutrino production from pulsar wind nebulae assuming that all TeV gamma-ray emission is from pions created in proton-proton interactions.
For ESTES, the $Fermi$-LAT $\pi^0$ and the supernova remnant analyses returned similar p-values, which are both consistent with the background only hypothesis.
One of the next steps in Galactic neutrino astronomy will be disentangling diffuse Galactic plane neutrinos from neutrinos produced at Galactic accelerators.

A near-future solution that can better resolve the Galactic plane is to combine the multiple event selection streams within IceCube to take advantage of all the astrophysical neutrino candidate events recorded by the detector.
An analysis has been proposed~\citep{IceCube:NED2023}, combining the ESTES starting tracks with the DNN cascade events~\citep{IceCube:DNNCSci} and the Earth through-going track events from a tracks sample~\citep{IceCube:NGC1068Sci}.
The tracks sample was used to search for galactic neutrino excesses using a binned style analysis which rejected the isotropic null hypothesis at the 2.7$\sigma$ level~\citep{IceCube:2023NTGPICRC}.
However, this tracks sample sample does not have a view of the central Galactic region where most emission is expected nor does it benefit from the atmospheric-neutrino veto effect that ESTES and DNN Cascades experience.
Using the combined sample, IceCube can further increase the significance of its detection of Galactic plane neutrinos and potentially provide some insight on the exact origin and energy spectrum of Galactic plane neutrinos.
The combined sample can also look for neutrino emission from extra-galactic sources as more information is learned from the recent NGC 1068 discovery~\citep{IceCube:NGC1068Sci}.
Many x-ray bright Seyfert galaxies like NGC 1068 are found in the southern sky and the combination of the pointing resolution of tracks and event rate of the cascades could help further identify these potential neutrino sources~\citep{IceCube:ESTESSeyferts2023}.

\section{Conclusions} \label{sec:Conclusion}
The Enhanced Starting Track Event Selection (ESTES) is a new selection of IceCube starting-track events which provides a unique sample of astrophysical muon neutrino candidates, focusing on neutrinos with energies between 1 and 500 TeV.
The advantages of this event selection come from the improved properties of energy resolution, pointing resolution, and background rejection for starting-track events.
In particular, selecting for starting events removes not only atmospheric muons created by cosmic-ray air showers, but also atmospheric neutrinos which are accompanied into the detector by muons from the same shower.
This increases the purity of the astrophysical neutrino sample in the southern equatorial sky ($\delta < -15^{\circ}$), resulting in an atmospheric background rate approximately 1 order of magnitude larger than the astrophysical neutrino rate compared to 4 orders of magnitude in the case of previous track selections~\citep{IceCube:10yrPSTracks, IceCube:7yrPSTracks}.
The ESTES analyses improve upon IceCube's sensitivity to southern-sky neutrino sources for sources with spectral indices less than -2.0 or energy cutoffs.
Although no new sources of astrophysical neutrinos were identified in the analysis of the ESTES data, we provide some of the strongest constraints for potential neutrino sources in the southern sky, including a 90\% statistics-only upper limit of 85\% of the KRA$_{\gamma}$ 5 PeV model prediction for the diffuse Galactic plane neutrino emission.

The ESTES sample also has other future uses for both multi-messenger astronomy and particle physics.
IceCube sends neutrino alerts to the astrophysical community for follow-up for events with a high likelihood of being astrophysical in origin~\citep{IceCube:2023IceCat}.
Due to the atmospheric-neutrino background, most of these events must have energies of around 100 TeV or greater to be sent as an alert.
However, with ESTES, this energy threshold can be lowered due to the atmospheric-neutrino veto in the southern sky and the improved energy resolution of starting tracks over incoming events.
By sending out starting-track alert events using ESTES, IceCube can provide neutrino alert events with a 50\% or greater likelihood of astrophysical origin with energies in the tens of TeV.
These events are of particular interest for real-time follow-up as transient sources located within our own Galaxy or nearby galaxies, such as tidal disruption events, may not have the power to produce 100 TeV neutrinos due to size limitations.
Nearby transients provide a good opportunity for gamma-ray follow-up as the gamma-rays are less likely to be attenuated.
The ESTES based realtime alerts would add an estimated 5.4 ``Gold'' events per year to the current IceCube alerts~\citep{IceCube:ESTReSICRC2023}.

Finally, for particle physics, the ESTES starting-track events provide a sample of over 10,000 muon neutrino CC interactions from 1 TeV to 500 TeV contained inside the detector geometry.
These events can be used to measure the inelasticity--the ratio of energy transferred to the hadronic shower to the initial neutrino energy--which gives insight into the atmospheric neutrino-to-anti-neutrino ratio.
Most future proposed cubic kilometer neutrino experiments will be located in the northern hemisphere and should therefore have a good resolution of the southern sky and Galactic center.
However, for IceCube-Gen2, developing a starting-track event selection has been considered a goal in order to generate a sample of muon neutrino events with atmospheric-neutrino suppression for studying the properties of the astrophysical neutrino flux below 100 TeV.

\begin{acknowledgments}
The IceCube collaboration recognizes the significant contributions to this work by Sarah Mancina and Manuel Silva. 
The authors gratefully acknowledge the support from the following agencies and institutions:
USA {\textendash} U.S. National Science Foundation-Office of Polar Programs,
U.S. National Science Foundation-Physics Division,
U.S. National Science Foundation-EPSCoR,
U.S. National Science Foundation-Office of Advanced Cyberinfrastructure,
Wisconsin Alumni Research Foundation,
Center for High Throughput Computing (CHTC) at the University of Wisconsin{\textendash}Madison,
Open Science Grid (OSG),
Partnership to Advance Throughput Computing (PATh),
Advanced Cyberinfrastructure Coordination Ecosystem: Services {\&} Support (ACCESS),
Frontera computing project at the Texas Advanced Computing Center,
U.S. Department of Energy-National Energy Research Scientific Computing Center,
Particle astrophysics research computing center at the University of Maryland,
Institute for Cyber-Enabled Research at Michigan State University,
Astroparticle physics computational facility at Marquette University,
NVIDIA Corporation,
and Google Cloud Platform;
Belgium {\textendash} Funds for Scientific Research (FRS-FNRS and FWO),
FWO Odysseus and Big Science programmes,
and Belgian Federal Science Policy Office (Belspo);
Germany {\textendash} Bundesministerium f{\"u}r Bildung und Forschung (BMBF),
Deutsche Forschungsgemeinschaft (DFG),
Helmholtz Alliance for Astroparticle Physics (HAP),
Initiative and Networking Fund of the Helmholtz Association,
Deutsches Elektronen Synchrotron (DESY),
and High Performance Computing cluster of the RWTH Aachen;
Sweden {\textendash} Swedish Research Council,
Swedish Polar Research Secretariat,
Swedish National Infrastructure for Computing (SNIC),
and Knut and Alice Wallenberg Foundation;
European Union {\textendash} EGI Advanced Computing for research;
Australia {\textendash} Australian Research Council;
Canada {\textendash} Natural Sciences and Engineering Research Council of Canada,
Calcul Qu{\'e}bec, Compute Ontario, Canada Foundation for Innovation, WestGrid, and Digital Research Alliance of Canada;
Denmark {\textendash} Villum Fonden, Carlsberg Foundation, and European Commission;
New Zealand {\textendash} Marsden Fund;
Japan {\textendash} Japan Society for Promotion of Science (JSPS)
and Institute for Global Prominent Research (IGPR) of Chiba University;
Korea {\textendash} National Research Foundation of Korea (NRF);
Switzerland {\textendash} Swiss National Science Foundation (SNSF).

\end{acknowledgments}
%

\vspace{5mm}
\facilities{IceCube}


\software{Astropy~\citep{astropy:2013, astropy:2018, astropy:2022},
          Healpy~\citep{Zonca2019, 2005ApJ:622:759G},
          LightGBM~\citep{lightgbm}
          }

\appendix

\section{Full Source-List Results and Upper Limits} \label{App:SrcListTables}

Here we present the individual results for each source and catalog tested by the gamma-ray informed source-list analyses. 
Tables~\ref{tab:GPStackingResults_SNRPWNUID} and~\ref{tab:GPStackingResults_BIN} list the sources tested in each stacking catalog along with the analysis results for the full catalog and the sources that were tested individually.
Table~\ref{tab:SSCatalogResults} lists all the single sources tested ordered by the log of the pre-trial p-value returned by the analysis.

\begin{table*}[tbh!]
    \centering
    \arraybackslash
    
    \begin{tabular}{|rccccc|cccc|}
    \hline
    Source Name & $\delta$ ($^{\circ}$) & $\alpha$ ($^{\circ}$) & $\phi_{\nu+\bar{\nu}}^{1 \mathrm{TeV}}$ & $\gamma$ & E$_{\mathrm{cut}}$ & TS & $-\log_{10}(p_{\mathrm{pre}})$ & 90\% Sens & 90\% UL  \\
    \hline
\textbf{SNR Catalog} & & & & & & 3.77 & 1.40 & 220\% & 600\% \\
Vela Junior & -46.33 & 133.00 & 1.4$ \times 10^{-11}$ & -1.81 & 3.34 &  0.00 & 0.46 &  1000\% & 1000\% \\
RX J1713.7-3946 & -39.77 & 258.36 & 9.4$ \times 10^{-12}$ & -2.06 & 6.45 &  0.00 & 0.61 &  960\% & 1000\% \\
HESS J1614-518 & -51.82 & 243.56 & 3.1$ \times 10^{-12}$ & -2.42 & -- &  & & & \\
HESS J1457-593 & -59.07 & 223.70 & 2.3$ \times 10^{-12}$ & -2.52 & -- &  & & & \\
SNR G323.7-01.0 & -57.20 & 233.63 & 1.1$ \times 10^{-12}$ & -2.51 & -- &  & & & \\
HESS J1731-347 & -34.71 & 262.98 & 1.1$ \times 10^{-12}$ & -2.32 & -- &  & & & \\
Gamma Cygni & +40.52 & 305.27 & 1.3$ \times 10^{-12}$ & -3.02 & -- &  & & & \\
RCW 86 & -62.65 & 220.12 & 1.1$ \times 10^{-12}$ & -1.60 & 1.75 &  & & & \\
HESS J1912+101 & +10.19 & 288.33 & 1.3$ \times 10^{-12}$ & -2.56 & -- &  & & & \\
HESS J1745-303 & -30.20 & 266.30 & 8.7$ \times 10^{-13}$ & -2.71 & -- &  & & & \\
Cassiopeia A & +58.81 & 350.85 & 4.1$ \times 10^{-13}$ & -2.61 & -- &  & & & \\
CTB 37A & -38.55 & 258.64 & 3.5$ \times 10^{-13}$ & -2.30 & -- &  & & & \\
        \hline
    \end{tabular}

\smallskip
    \begin{tabular}{|rccccc|cccc|}
    \hline
    Source Name & $\delta$ ($^{\circ}$) & $\alpha$ ($^{\circ}$) & $\phi_{\nu+\bar{\nu}}^{1 \mathrm{TeV}}$ & $\gamma$ & E$_{\mathrm{cut}}$ & TS & $-\log_{10}(p_{\mathrm{pre}})$ & 90\% Sens & 90\% UL  \\
    \hline
\textbf{PWN Catalog} & & & & & & 0.67 & 0.54 & 54\% & 77\% \\
Vela X & -45.19 & 128.29 & 1.0$ \times 10^{-11}$ & -1.32 & 7.00 &  0.00 & 0.53 &  170\% & 170\% \\
Crab Nebula & +22.01 & 83.63 & 1.2$ \times 10^{-11}$ & -2.39 & 7.15 &  0.00 & 0.39 &  780\% & 820\% \\
HESS J1708-443 & -44.30 & 257.00 & 2.1$ \times 10^{-12}$ & -2.00 & -- &  0.00 & 0.57 &  250\% & 250\% \\
HESS J1825-137 & -13.58 & 276.55 & 8.1$ \times 10^{-12}$ & -2.26 & 12.40 &  & & & \\
HESS J1632-478 & -47.87 & 248.01 & 2.4$ \times 10^{-12}$ & -2.12 & -- &  & & & \\
MSH 15-52 & -59.16 & 228.53 & 2.4$ \times 10^{-12}$ & -2.27 & -- &  & & & \\
HESS J1813-178 & -17.85 & 273.36 & 1.3$ \times 10^{-12}$ & -2.09 & -- &  & & & \\
HESS J1303-631 & -63.20 & 195.75 & 3.1$ \times 10^{-12}$ & -1.50 & 3.85 &  & & & \\
HESS J1616-508 & -50.91 & 244.06 & 2.6$ \times 10^{-12}$ & -2.35 & -- &  & & & \\
HESS J1420-607 & -60.98 & 214.69 & 1.5$ \times 10^{-12}$ & -2.20 & -- &  & & & \\
HESS J1837-069 & -6.93 & 279.43 & 2.1$ \times 10^{-12}$ & -2.27 & -- &  & & & \\
HESS J1026-582 & -58.29 & 157.17 & 5.2$ \times 10^{-13}$ & -1.94 & -- &  & & & \\
        \hline
    \end{tabular}

\smallskip
    \begin{tabular}{|rccccc|cccc|}
    \hline
    Source Name & $\delta$ ($^{\circ}$) & $\alpha$ ($^{\circ}$) & $\phi_{\nu+\bar{\nu}}^{1 \mathrm{TeV}}$ & $\gamma$ & E$_{\mathrm{cut}}$ & TS & $-\log_{10}(p_{\mathrm{pre}})$ & 90\% Sens & 90\% UL  \\
    \hline
\textbf{UID Catalog} & & & & & & 0.29 & 0.51 & 75\% & 100\% \\
MGRO J1908+06 & +6.32 & 286.91 & 1.5$ \times 10^{-11}$ & -2.70 & -- &  0.00 & 0.36 &  330\% & 350\% \\
Westerlund 1 & -45.80 & 251.50 & 3.9$ \times 10^{-12}$ & -2.19 & -- &  0.00 & 0.49 &  250\% & 260\% \\
HESS J1702-420 & -42.02 & 255.68 & 2.3$ \times 10^{-12}$ & -2.09 & -- &  & & & \\
2HWC J1814-173 & -17.31 & 273.52 & 4.7$ \times 10^{-12}$ & -2.61 & -- &  & & & \\
HESS J1841-055 & -5.55 & 280.23 & 5.6$ \times 10^{-12}$ & -2.47 & -- &  & & & \\
2HWC J1819-150 & -15.06 & 274.83 & 4.3$ \times 10^{-12}$ & -2.88 & -- &  & & & \\
HESS J1804-216 & -21.73 & 271.12 & 2.7$ \times 10^{-12}$ & -2.69 & -- &  & & & \\
HESS J1809-193 & -19.30 & 272.63 & 2.9$ \times 10^{-12}$ & -2.38 & -- &  & & & \\
HESS J1843-033 & -3.30 & 280.75 & 1.6$ \times 10^{-12}$ & -2.15 & -- &  & & & \\
TeV J2032+4130 & +41.51 & 307.93 & 2.9$ \times 10^{-12}$ & -2.52 & -- &  & & & \\
HESS J1708-410 & -41.09 & 257.10 & 1.9$ \times 10^{-12}$ & -2.58 & -- &  & & & \\
HESS J1857+026 & +2.67 & 284.30 & 2.1$ \times 10^{-12}$ & -2.57 & -- &  & & & \\
        \hline
    \end{tabular}

    \caption{Results from searches for Galactic plane point-like sources of neutrinos. The results for the stacking catalog searches for each source classification are given in the top lines of each table: supernova remnants (SNR), pulsar wind nebulae (PWN), and unidentified TeV Galactic plane objects (UID). Below the stacking results are the individual sources included in the stacking catalogs with the flux assumptions used to model each source given. The results for the seven individual Galactic plane sources are also given in line with the source information. The flux reported on the left, $\phi_{\nu+\bar{\nu}}^{1 \mathrm{TeV}}$ , is the neutrino and anti-neutrino per-flavor flux at 1 TeV in units of TeV$^{-1}$ cm$^{-2}$ s$^{-1}$. Some source models include an exponential energy cutoff which is given in units of TeV. The sensitivity and statistics-only upper limits are given as a percentage of the injected model flux given by the model parameters.
    }
    \label{tab:GPStackingResults_SNRPWNUID}
\end{table*}

\begin{table*}[tbh!]
    \centering
    \begin{tabular}{|rcc|cccc|}
    \hline
    Source Name & $\delta$ ($^{\circ}$) & $\alpha$ ($^{\circ}$) & TS & $-\log_{10}(p_{\mathrm{pre}})$ & $\phi^{\gamma = -2.0}_{1\mathrm{TeV}}$ 90\% UL & $\phi^{\gamma = -3.0}_{1\mathrm{TeV}}$ 90\% UL   \\
    \hline
\textbf{BIN Catalog} & & & 0.00 & 0.31 & 3.0$ \times 10^{-12}$ & 9.2$ \times 10^{-13}$ \\
LS I +61 303 & +61.23 & 40.13 &  & & &  \\
LS 5039 & -14.84 & 276.56 &  & & &  \\
PSR B1259-63 & -63.84 & 195.70 &  & & &  \\
LMC P3 & -67.59 & 84.00 &  & & &  \\
HESS J0632+057 & +5.80 & 98.25 &  & & &  \\
Eta Carinae & -59.68 & 161.27 &  & & &  \\
PSR J2032+4127 & +41.46 & 308.05 &  & & &  \\
HESS J1018-589 A & -58.93 & 154.75 &  & & &  \\
HESS J1832-093 & -9.37 & 278.19 &  & & &  \\
        \hline
    \end{tabular}
    \caption{Results from the TeV binary stacking analysis. This catalog does not include models due to the variability of the gamma-ray flux from binary systems. The sources included in the catalog are listed with their position under the catalog name. The upper limit fluxes from the analysis are given assuming equal emission from each source and a spectral index of $\gamma = -2.0$ and $\gamma = -3.0$. The statistics-only upper limit results are given as the neutrino and anti-neutrino per-flavor flux in units of TeV$^{-1}$ cm$^{-2}$ s$^{-1}$.}
    \label{tab:GPStackingResults_BIN}
\end{table*}

\centering
\begin{longtable}{rcccccccc}
\caption{Results for each source from the gamma-ray influenced single source list given in order of pre-trial p-value. The classifications of each source are given by the catalogs used~\citep{Fermi-LAT:Catalog,GammaCat,TeVCat}. The best fit values for the number of signal neutrinos, $\hat{n}_s$, are given for all sources, and the best fit values for the spectral index, $\hat{\gamma}$, are given for the top 20 sources in the list. When maximizing the likelihood, the spectral index allowed to float between -1.0 and -4.0. The 90\% statistics-only upper limits are reported assuming two different spectral indicies, $\gamma=-2.0$ and $\gamma=-3.0$, and are given as the neutrino and anti-neutrino per-flavor flux at 1 TeV in units of TeV$^{-1}$ cm$^{-2}$ s$^{-1}$.}\label{tab:SSCatalogResults}\\
\toprule
            Name &  Type &    $\delta$ $(^{\circ}$) &  $-\log_{10}(\mathrm{p}_{\mathrm{pre}})$ &  TS & $\hat{n}_s$ & $\hat{\gamma}$ & $\phi_{\mathrm{1 TeV}}^{\gamma=-2.0}$ 90\% UL &  $\phi_{\mathrm{1 TeV}}^{\gamma=-3.0}$ 90\% UL \\
            \hline
\midrule
\endfirsthead

\toprule
            Name &  Type &    $\delta$ $(^{\circ}$) &  $-\log_{10}(\mathrm{p}_{\mathrm{pre}})$ &  TS & $\hat{n}_s$ & $\hat{\gamma}$ & $\phi_{\mathrm{1 TeV}}^{\gamma=-2.0}$ 90\% UL &  $\phi_{\mathrm{1 TeV}}^{\gamma=-3.0}$ 90\% UL \\
            \hline
\midrule
\endhead
\midrule
\multicolumn{7}{r}{{Continued on next page}} \\
\midrule
\endfoot

\bottomrule
\endlastfoot
1ES 0647+250 & BLL & +25.05 & 3.39 & 10.6 & 11.3 & -4.0 & 1.7$ \times 10^{-11}$ & 3.9$ \times 10^{-10}$ \\
1H 1720+117 & BLL & +11.87 & 2.32 & 6.1 & 1.6 & -1.6 & 1.3$ \times 10^{-11}$ & 3.3$ \times 10^{-10}$ \\
PKS 0727-11 & FSRQ & -11.69 & 2.31 & 5.8 & 6.5 & -3.6 & 1.3$ \times 10^{-11}$ & 4.2$ \times 10^{-10}$ \\
1ES 1959+650 & BLL & +65.15 & 1.89 & 4.8 & 5.0 & -4.0 & 1.7$ \times 10^{-11}$ & 2.6$ \times 10^{-10}$ \\
OT 081 & BLL & +9.65 & 1.78 & 4.0 & 6.5 & -3.7 & 1.1$ \times 10^{-11}$ & 2.9$ \times 10^{-10}$ \\
SMC & GAL & -72.75 & 1.52 & 2.8 & 1.6 & -2.1 & 7.2$ \times 10^{-12}$ & 3.2$ \times 10^{-10}$ \\
B3 0133+388 & BLL & +39.10 & 1.47 & 2.9 & 4.3 & -3.2 & 1.1$ \times 10^{-11}$ & 2.2$ \times 10^{-10}$ \\
B3 1343+451 & FSRQ & +44.88 & 1.45 & 3.6 & 1.8 & -2.2 & 1.2$ \times 10^{-11}$ & 2.1$ \times 10^{-10}$ \\
PKS 0332-403 & BLL & -40.15 & 1.35 & 2.0 & 1.5 & -3.1 & 7.3$ \times 10^{-12}$ & 2.2$ \times 10^{-10}$ \\
NGC 1068 & SBG & -0.01 & 1.28 & 3.3 & 8.1 & -4.0 & 9.6$ \times 10^{-12}$ & 2.5$ \times 10^{-10}$ \\
PKS 2247-131 & BCU & -12.85 & 1.21 & 1.9 & 3.3 & -3.2 & 9.6$ \times 10^{-12}$ & 2.8$ \times 10^{-10}$ \\
OJ 287 & BLL & +20.12 & 1.12 & 1.6 & 3.7 & -3.9 & 9.5$ \times 10^{-12}$ & 2.0$ \times 10^{-10}$ \\
PKS 0700-661 & BLL & -66.18 & 0.99 & 1.4 & 1.1 & -2.5 & 6.1$ \times 10^{-12}$ & 2.3$ \times 10^{-10}$ \\
B3 0609+413 & BLL & +41.37 & 0.97 & 1.3 & 2.1 & -2.7 & 9.3$ \times 10^{-12}$ & 1.7$ \times 10^{-10}$ \\
TXS 0506+056 & BLL & +5.70 & 0.92 & 1.3 & 5.2 & -4.0 & 8.0$ \times 10^{-12}$ & 2.0$ \times 10^{-10}$ \\
M 82 & SBG & +69.67 & 0.91 & 1.2 & 3.0 & -3.9 & 1.2$ \times 10^{-11}$ & 1.8$ \times 10^{-10}$ \\
Cen B & RDG & -60.45 & 0.83 & 1.1 & 1.9 & -3.5 & 5.7$ \times 10^{-12}$ & 1.9$ \times 10^{-10}$ \\
PKS 0521-36 & AGN & -36.47 & 0.80 & 0.4 & 0.7 & -4.0 & 6.1$ \times 10^{-12}$ & 1.7$ \times 10^{-10}$ \\
PMN J1918-4111 & BLL & -41.19 & 0.80 & 0.3 & 0.7 & -2.5 & 6.1$ \times 10^{-12}$ & 1.7$ \times 10^{-10}$ \\
PMN J0531-4827 & BLL & -48.46 & 0.76 & 1.2 & 1.3 & -4.0 & 5.8$ \times 10^{-12}$ & 1.7$ \times 10^{-10}$ \\
NGC 3424 & SBG & +32.89 & 0.76 & 0.9 & 2.4 & - & 8.1$ \times 10^{-12}$ & 1.6$ \times 10^{-10}$ \\
PKS 0208-512 & FSRQ & -51.02 & 0.75 & 2.0 & 2.1 & - & 6.2$ \times 10^{-12}$ & 1.8$ \times 10^{-10}$ \\
PKS 0235+164 & BLL & +16.62 & 0.75 & 0.7 & 3.0 & - & 7.8$ \times 10^{-12}$ & 1.7$ \times 10^{-10}$ \\
PKS 0447-439 & BLL & -43.84 & 0.70 & 0.1 & 0.4 & - & 5.6$ \times 10^{-12}$ & 1.5$ \times 10^{-10}$ \\
PKS 0502+049 & FSRQ & +5.00 & 0.69 & 0.7 & 3.3 & - & 7.0$ \times 10^{-12}$ & 1.7$ \times 10^{-10}$ \\
PKS 0735+17 & BLL & +17.71 & 0.68 & 0.5 & 1.2 & - & 7.6$ \times 10^{-12}$ & 1.6$ \times 10^{-10}$ \\
OJ 014 & BLL & +1.78 & 0.66 & 1.0 & 4.7 & - & 6.8$ \times 10^{-12}$ & 1.6$ \times 10^{-10}$ \\
PKS 2142-75 & FSRQ & -75.60 & 0.66 & 0.0 & 0.0 & - & 5.0$ \times 10^{-12}$ & 2.0$ \times 10^{-10}$ \\
PKS 1502+106 & FSRQ & +10.50 & 0.66 & 0.4 & 2.0 & - & 7.2$ \times 10^{-12}$ & 1.6$ \times 10^{-10}$ \\
PKS 1424-41 & FSRQ & -42.11 & 0.61 & 0.0 & 0.0 & - & 5.3$ \times 10^{-12}$ & 1.4$ \times 10^{-10}$ \\
RX J1713.7-3946 & SNR & -39.77 & 0.61 & 0.0 & 0.0 & - & 5.3$ \times 10^{-12}$ & 1.4$ \times 10^{-10}$ \\
PKS 1440-389 & BLL & -39.15 & 0.61 & 0.0 & 0.0 & - & 5.5$ \times 10^{-12}$ & 1.4$ \times 10^{-10}$ \\
PKS 0426-380 & BLL & -37.94 & 0.61 & 0.0 & 0.0 & - & 5.4$ \times 10^{-12}$ & 1.4$ \times 10^{-10}$ \\
Cen A & RDG & -43.02 & 0.61 & 0.0 & 0.0 & - & 5.2$ \times 10^{-12}$ & 1.4$ \times 10^{-10}$ \\
MH 2136-428 & BLL & -42.59 & 0.61 & 0.0 & 0.0 & - & 5.3$ \times 10^{-12}$ & 1.4$ \times 10^{-10}$ \\
4C +55.17 & FSRQ & +55.38 & 0.60 & 0.9 & 0.7 & - & 8.3$ \times 10^{-12}$ & 1.3$ \times 10^{-10}$ \\
LMC & GAL & -68.75 & 0.59 & 0.0 & 0.0 & - & 4.8$ \times 10^{-12}$ & 1.7$ \times 10^{-10}$ \\
1H 1013+498 & BLL & +49.43 & 0.58 & 0.8 & 0.7 & - & 7.9$ \times 10^{-12}$ & 1.3$ \times 10^{-10}$ \\
4C +01.28 & BLL & +1.56 & 0.58 & 0.7 & 2.7 & - & 6.6$ \times 10^{-12}$ & 1.5$ \times 10^{-10}$ \\
PKS 0537-441 & BLL & -44.09 & 0.58 & 0.0 & 0.0 & - & 5.2$ \times 10^{-12}$ & 1.4$ \times 10^{-10}$ \\
PKS 0301-243 & BLL & -24.12 & 0.58 & 0.2 & 0.6 & - & 6.0$ \times 10^{-12}$ & 1.6$ \times 10^{-10}$ \\
3C 279 & FSRQ & -5.79 & 0.58 & 0.3 & 1.4 & - & 6.4$ \times 10^{-12}$ & 1.6$ \times 10^{-10}$ \\
HESS J1708-443 & PWN & -44.30 & 0.57 & 0.0 & 0.0 & - & 5.2$ \times 10^{-12}$ & 1.4$ \times 10^{-10}$ \\
PMN J1610-6649 & BLL & -66.81 & 0.57 & 0.0 & 0.0 & - & 4.7$ \times 10^{-12}$ & 1.6$ \times 10^{-10}$ \\
1H 1914-194 & BLL & -19.36 & 0.54 & 0.1 & 0.7 & - & 6.2$ \times 10^{-12}$ & 1.6$ \times 10^{-10}$ \\
4C +01.02 & FSRQ & +1.58 & 0.54 & 0.6 & 3.1 & - & 6.2$ \times 10^{-12}$ & 1.5$ \times 10^{-10}$ \\
Vela X & PWN & -45.19 & 0.53 & 0.0 & 0.0 & - & 5.0$ \times 10^{-12}$ & 1.3$ \times 10^{-10}$ \\
Mkn 421 & BLL & +38.21 & 0.52 & 0.1 & 1.0 & - & 6.8$ \times 10^{-12}$ & 1.2$ \times 10^{-10}$ \\
Arp 299 & SBG & +58.52 & 0.51 & 0.4 & 1.8 & - & 7.7$ \times 10^{-12}$ & 1.2$ \times 10^{-10}$ \\
Westerlund 1 & UID & -45.80 & 0.49 & 0.0 & 0.0 & - & 4.8$ \times 10^{-12}$ & 1.3$ \times 10^{-10}$ \\
Ton 599 & FSRQ & +29.24 & 0.49 & 0.1 & 0.7 & - & 6.6$ \times 10^{-12}$ & 1.2$ \times 10^{-10}$ \\
NGC 1275 & RDG & +41.51 & 0.48 & 0.1 & 0.9 & - & 6.6$ \times 10^{-12}$ & 1.2$ \times 10^{-10}$ \\
PKS 1936-623 & BLL & -62.18 & 0.48 & 0.0 & 0.0 & - & 4.4$ \times 10^{-12}$ & 1.4$ \times 10^{-10}$ \\
BL Lac & BLL & +42.28 & 0.48 & 0.1 & 0.8 & - & 6.7$ \times 10^{-12}$ & 1.1$ \times 10^{-10}$ \\
PKS 2155-304 & BLL & -30.23 & 0.47 & 0.0 & 0.0 & - & 5.2$ \times 10^{-12}$ & 1.4$ \times 10^{-10}$ \\
PKS 1101-536 & BLL & -53.96 & 0.47 & 0.6 & 1.3 & - & 4.9$ \times 10^{-12}$ & 1.4$ \times 10^{-10}$ \\
Vela Junior & SNR & -46.33 & 0.46 & 0.0 & 0.0 & - & 4.8$ \times 10^{-12}$ & 1.2$ \times 10^{-10}$ \\
B2 0218+357 & FSRQ & +35.94 & 0.45 & 0.1 & 0.2 & - & 6.4$ \times 10^{-12}$ & 1.1$ \times 10^{-10}$ \\
Galactic Centre & BCU & -29.00 & 0.45 & 0.0 & 0.0 & - & 5.3$ \times 10^{-12}$ & 1.4$ \times 10^{-10}$ \\
PKS 1124-186 & FSRQ & -18.96 & 0.44 & 0.0 & 0.0 & - & 5.7$ \times 10^{-12}$ & 1.4$ \times 10^{-10}$ \\
PKS 1830-211 & FSRQ & -21.06 & 0.44 & 0.0 & 0.0 & - & 5.4$ \times 10^{-12}$ & 1.4$ \times 10^{-10}$ \\
PKS 0823-223 & BLL & -22.51 & 0.44 & 0.0 & 0.0 & - & 5.2$ \times 10^{-12}$ & 1.3$ \times 10^{-10}$ \\
PKS 0454-234 & FSRQ & -23.41 & 0.43 & 0.0 & 0.0 & - & 5.2$ \times 10^{-12}$ & 1.4$ \times 10^{-10}$ \\
TXS 0628-240 & BLL & -24.11 & 0.43 & 0.0 & 0.0 & - & 5.2$ \times 10^{-12}$ & 1.3$ \times 10^{-10}$ \\
PKS 0118-272 & BLL & -27.02 & 0.43 & 0.0 & 0.0 & - & 5.1$ \times 10^{-12}$ & 1.3$ \times 10^{-10}$ \\
NGC 253 & SBG & -25.29 & 0.43 & 0.0 & 0.0 & - & 5.2$ \times 10^{-12}$ & 1.4$ \times 10^{-10}$ \\
S5 0716+71 & BLL & +71.34 & 0.43 & 0.0 & 0.4 & - & 8.8$ \times 10^{-12}$ & 1.2$ \times 10^{-10}$ \\
PKS 1244-255 & FSRQ & -25.80 & 0.43 & 0.0 & 0.0 & - & 5.2$ \times 10^{-12}$ & 1.3$ \times 10^{-10}$ \\
PMN J2345-1555 & FSRQ & -15.92 & 0.43 & 0.0 & 0.0 & - & 5.9$ \times 10^{-12}$ & 1.4$ \times 10^{-10}$ \\
AP Librae & BLL & -24.37 & 0.43 & 0.0 & 0.0 & - & 5.1$ \times 10^{-12}$ & 1.3$ \times 10^{-10}$ \\
PKS 0805-07 & FSRQ & -7.86 & 0.43 & 0.0 & 0.2 & - & 5.6$ \times 10^{-12}$ & 1.4$ \times 10^{-10}$ \\
PKS 1510-089 & FSRQ & -9.11 & 0.42 & 0.0 & 0.0 & - & 5.6$ \times 10^{-12}$ & 1.5$ \times 10^{-10}$ \\
PKS 0048-09 & BLL & -9.49 & 0.42 & 0.0 & 0.0 & - & 5.7$ \times 10^{-12}$ & 1.5$ \times 10^{-10}$ \\
PKS 2233-148 & BLL & -14.56 & 0.42 & 0.0 & 0.0 & - & 5.7$ \times 10^{-12}$ & 1.5$ \times 10^{-10}$ \\
PKS 1730-13 & FSRQ & -13.09 & 0.42 & 0.0 & 0.0 & - & 5.8$ \times 10^{-12}$ & 1.5$ \times 10^{-10}$ \\
PKS 2023-07 & FSRQ & -7.59 & 0.41 & 0.0 & 0.0 & - & 5.2$ \times 10^{-12}$ & 1.3$ \times 10^{-10}$ \\
S2 0109+22 & BLL & +22.75 & 0.40 & 0.0 & 0.2 & - & 5.9$ \times 10^{-12}$ & 1.1$ \times 10^{-10}$ \\
NGC 5380 & GAL & +37.50 & 0.40 & 0.0 & 0.0 & - & 6.0$ \times 10^{-12}$ & 1.0$ \times 10^{-10}$ \\
RGB J2243+203 & BLL & +20.36 & 0.40 & 0.0 & 0.0 & - & 5.7$ \times 10^{-12}$ & 1.1$ \times 10^{-10}$ \\
4C +38.41 & FSRQ & +38.14 & 0.39 & 0.0 & 0.0 & - & 5.9$ \times 10^{-12}$ & 9.9$ \times 10^{-11}$ \\
4C +21.35 & FSRQ & +21.38 & 0.39 & 0.0 & 0.0 & - & 5.6$ \times 10^{-12}$ & 1.1$ \times 10^{-10}$ \\
Crab nebula & PWN & +22.01 & 0.39 & 0.0 & 0.0 & - & 5.6$ \times 10^{-12}$ & 1.0$ \times 10^{-10}$ \\
MG2 J201534+3710 & FSRQ & +37.18 & 0.39 & 0.0 & 0.0 & - & 5.9$ \times 10^{-12}$ & 1.0$ \times 10^{-10}$ \\
NGC 2146 & SBG & +78.33 & 0.39 & 0.0 & 0.0 & - & 8.3$ \times 10^{-12}$ & 1.1$ \times 10^{-10}$ \\
TXS 0518+211 & BLL & +21.21 & 0.39 & 0.0 & 0.0 & - & 5.5$ \times 10^{-12}$ & 1.1$ \times 10^{-10}$ \\
Mkn 501 & BLL & +39.76 & 0.39 & 0.0 & 0.0 & - & 5.9$ \times 10^{-12}$ & 1.0$ \times 10^{-10}$ \\
MG1 J021114+1051 & BLL & +10.86 & 0.39 & 0.0 & 0.0 & - & 5.3$ \times 10^{-12}$ & 1.1$ \times 10^{-10}$ \\
PG 1553+113 & BLL & +11.19 & 0.38 & 0.0 & 0.0 & - & 5.3$ \times 10^{-12}$ & 1.1$ \times 10^{-10}$ \\
PKS 1424+240 & BLL & +23.80 & 0.38 & 0.0 & 0.0 & - & 5.6$ \times 10^{-12}$ & 1.0$ \times 10^{-10}$ \\
Arp 220 & SBG & +23.53 & 0.38 & 0.0 & 0.0 & - & 5.5$ \times 10^{-12}$ & 1.0$ \times 10^{-10}$ \\
ON 246 & BLL & +25.30 & 0.38 & 0.0 & 0.0 & - & 5.6$ \times 10^{-12}$ & 1.0$ \times 10^{-10}$ \\
TXS 2241+406 & FSRQ & +40.96 & 0.38 & 0.0 & 0.1 & - & 6.0$ \times 10^{-12}$ & 1.0$ \times 10^{-10}$ \\
CTA 102 & FSRQ & +11.73 & 0.38 & 0.0 & 0.0 & - & 5.3$ \times 10^{-12}$ & 1.1$ \times 10^{-10}$ \\
M 31 & GAL & +41.24 & 0.37 & 0.0 & 0.0 & - & 5.7$ \times 10^{-12}$ & 9.6$ \times 10^{-11}$ \\
TXS 0141+268 & BLL & +27.09 & 0.37 & 0.0 & 0.0 & - & 5.5$ \times 10^{-12}$ & 1.0$ \times 10^{-10}$ \\
IC 678 & GAL & +6.63 & 0.37 & 0.0 & 0.0 & - & 4.8$ \times 10^{-12}$ & 1.1$ \times 10^{-10}$ \\
B2 1215+30 & BLL & +30.12 & 0.36 & 0.0 & 0.0 & - & 5.4$ \times 10^{-12}$ & 1.0$ \times 10^{-10}$ \\
MGRO J1908+06 & UID & +6.32 & 0.36 & 0.0 & 0.0 & - & 4.8$ \times 10^{-12}$ & 1.0$ \times 10^{-10}$ \\
4C +28.07 & FSRQ & +28.80 & 0.36 & 0.0 & 0.0 & - & 5.4$ \times 10^{-12}$ & 1.0$ \times 10^{-10}$ \\
S4 0814+42 & BLL & +42.38 & 0.36 & 0.0 & 0.0 & - & 5.5$ \times 10^{-12}$ & 8.9$ \times 10^{-11}$ \\
3C 454.3 & FSRQ & +16.15 & 0.35 & 0.0 & 0.0 & - & 5.2$ \times 10^{-12}$ & 1.0$ \times 10^{-10}$ \\
B2 1520+31 & FSRQ & +31.74 & 0.35 & 0.0 & 0.0 & - & 5.3$ \times 10^{-12}$ & 9.6$ \times 10^{-11}$ \\
3C 66A & BLL & +43.04 & 0.35 & 0.0 & 0.0 & - & 5.4$ \times 10^{-12}$ & 8.8$ \times 10^{-11}$ \\
PMN J1329-5608 & BLL & -56.12 & 0.34 & 0.0 & 0.0 & - & 4.0$ \times 10^{-12}$ & 1.1$ \times 10^{-10}$ \\
PMN J1603-4904 & BLL & -49.06 & 0.33 & 0.0 & 0.0 & - & 4.1$ \times 10^{-12}$ & 1.1$ \times 10^{-10}$ \\
NGC 4945 & SBG & -49.47 & 0.32 & 0.0 & 0.0 & - & 4.1$ \times 10^{-12}$ & 1.1$ \times 10^{-10}$ \\
PKS 0336-01 & FSRQ & -1.78 & 0.32 & 0.0 & 0.7 & - & 4.8$ \times 10^{-12}$ & 1.1$ \times 10^{-10}$ \\
GB6 J1542+6129 & BLL & +61.50 & 0.32 & 0.0 & 0.0 & - & 6.1$ \times 10^{-12}$ & 8.5$ \times 10^{-11}$ \\
PKS 2326-502 & FSRQ & -49.93 & 0.30 & 0.0 & 0.0 & - & 4.0$ \times 10^{-12}$ & 1.0$ \times 10^{-10}$ \\
PMN J1650-5044 & BLL & -50.75 & 0.29 & 0.0 & 0.0 & - & 3.9$ \times 10^{-12}$ & 1.0$ \times 10^{-10}$ \\
\end{longtable}

\bibliography{ESTES_NS}{}

\begin{thebibliography}{}
\expandafter\ifx\csname natexlab\endcsname\relax\def\natexlab#1{#1}\fi
\providecommand{\url}[1]{\href{#1}{#1}}
\providecommand{\dodoi}[1]{doi:~\href{http://doi.org/#1}{\nolinkurl{#1}}}
\providecommand{\doeprint}[1]{\href{http://ascl.net/#1}{\nolinkurl{http://ascl.net/#1}}}
\providecommand{\doarXiv}[1]{\href{https://arxiv.org/abs/#1}{\nolinkurl{https://arxiv.org/abs/#1}}}

\bibitem[{Aartsen {et~al.}(2013{\natexlab{a}})}]{IceCube:2013HESEDiscovery}
Aartsen, M.~G., {et~al.} 2013{\natexlab{a}}, Science, 342, 1242856, \dodoi{10.1126/science.1242856}

\bibitem[{Aartsen {et~al.}(2013{\natexlab{b}})}]{IceCube:3yrPSTracks2013}
---. 2013{\natexlab{b}}, Astrophys. J., 779, 132, \dodoi{10.1088/0004-637X/779/2/132}

\bibitem[{Aartsen {et~al.}(2015)}]{IceCube:MESE}
---. 2015, Phys. Rev., D91, 022001, \dodoi{10.1103/PhysRevD.91.022001}

\bibitem[{Aartsen {et~al.}(2017{\natexlab{a}})}]{IceCube:2017DetPaper}
---. 2017{\natexlab{a}}, Journal of Instrumentation, 12, P03012–P03012, \dodoi{10.1088/1748-0221/12/03/p03012}

\bibitem[{Aartsen {et~al.}(2017{\natexlab{b}})}]{IceCube:7yrPSTracks}
---. 2017{\natexlab{b}}, Astrophys. J., 835, 151, \dodoi{10.3847/1538-4357/835/2/151}

\bibitem[{Aartsen {et~al.}(2017{\natexlab{c}})}]{IceCube:PSTracks7YRGP}
---. 2017{\natexlab{c}}, Astrophys. J., 849, 67, \dodoi{10.3847/1538-4357/aa8dfb}

\bibitem[{Aartsen {et~al.}(2018{\natexlab{a}})}]{IceCube:TXSMMPaper}
---. 2018{\natexlab{a}}, Science, 361, eaat1378, \dodoi{10.1126/science.aat1378}

\bibitem[{Aartsen {et~al.}(2018{\natexlab{b}})}]{IceCube:TXSNeutFlare}
---. 2018{\natexlab{b}}, Science, 361, 147, \dodoi{10.1126/science.aat2890}

\bibitem[{Aartsen {et~al.}(2020{\natexlab{a}})}]{IceCube:10yrPSTracks}
---. 2020{\natexlab{a}}, Phys. Rev. Lett., 124, 051103, \dodoi{10.1103/PhysRevLett.124.051103}

\bibitem[{Aartsen {et~al.}(2020{\natexlab{b}})}]{IceCube:SCDiffuse}
---. 2020{\natexlab{b}}, Phys. Rev. Lett., 125, 121104, \dodoi{10.1103/PhysRevLett.125.121104}

\bibitem[{Abbasi {et~al.}(2021{\natexlab{a}})}]{IceCube:ESTESICRC2021}
Abbasi, R., {et~al.} 2021{\natexlab{a}}, PoS, ICRC2021, 1130, \dodoi{10.22323/1.395.1130}

\bibitem[{Abbasi {et~al.}(2021{\natexlab{b}})}]{IceCube:HESE7yr}
---. 2021{\natexlab{b}}, Phys. Rev. D, 104, 022002, \dodoi{10.1103/PhysRevD.104.022002}

\bibitem[{Abbasi {et~al.}(2022{\natexlab{a}})}]{IceCube:NGC1068Sci}
---. 2022{\natexlab{a}}, Science, 378, 538, \dodoi{10.1126/science.abg3395}

\bibitem[{Abbasi {et~al.}(2022{\natexlab{b}})}]{IceCube:2021DiffuseNumu}
---. 2022{\natexlab{b}}, Astrophys. J., 928, 50, \dodoi{10.3847/1538-4357/ac4d29}

\bibitem[{Abbasi {et~al.}(2023{\natexlab{a}})}]{IceCube:DNNCSci}
---. 2023{\natexlab{a}}, Science, 380, adc9818, \dodoi{10.1126/science.adc9818}

\bibitem[{Abbasi {et~al.}(2023{\natexlab{b}})}]{IceCube:2023IceCat}
---. 2023{\natexlab{b}}, Astrophys. J. Suppl., 269, 25, \dodoi{10.3847/1538-4365/acfa95}

\bibitem[{Abbasi {et~al.}(2024)}]{IceCube:ESTESPRD}
---. 2024, Phys. Rev. D, 110, 022001, \dodoi{10.1103/PhysRevD.110.022001}

\bibitem[{Abdollahi {et~al.}(2020)}]{Fermi-LAT:Catalog}
Abdollahi, S., {et~al.} 2020, Astrophys. J. Suppl., 247, 33, \dodoi{10.3847/1538-4365/ab6bcb}

\bibitem[{Ackermann {et~al.}(2012)}]{Fermi:FermiPi0Model}
Ackermann, M., {et~al.} 2012, Astrophys. J., 750, 3, \dodoi{10.1088/0004-637X/750/1/3}

\bibitem[{Albert {et~al.}(2017)}]{ANTARES:ANTARESFirstPS}
Albert, A., {et~al.} 2017, Phys. Rev. D, 96, 082001, \dodoi{10.1103/PhysRevD.96.082001}

\bibitem[{Albert {et~al.}(2018)}]{ANTARES:IceCube:GP}
---. 2018, The Astrophysical Journal, 868, L20, \dodoi{10.3847/2041-8213/aaeecf}

\bibitem[{Arg{\"u}elles {et~al.}(2018)Arg{\"u}elles, Palomares-Ruiz, Schneider, Wille, \& Yuan}]{Self_Veto:Arguelles}
Arg{\"u}elles, C.~A., Palomares-Ruiz, S., Schneider, A., Wille, L., \& Yuan, T. 2018, JCAP, 1807, 047, \dodoi{10.1088/1475-7516/2018/07/047}

\bibitem[{{Astropy Collaboration} {et~al.}(2013){Astropy Collaboration}, {Robitaille}, {Tollerud}, {Greenfield}, {Droettboom}, {Bray}, {Aldcroft}, {Davis}, {Ginsburg}, {Price-Whelan}, {Kerzendorf}, {Conley}, {Crighton}, {Barbary}, {Muna}, {Ferguson}, {Grollier}, {Parikh}, {Nair}, {Unther}, {Deil}, {Woillez}, {Conseil}, {Kramer}, {Turner}, {Singer}, {Fox}, {Weaver}, {Zabalza}, {Edwards}, {Azalee Bostroem}, {Burke}, {Casey}, {Crawford}, {Dencheva}, {Ely}, {Jenness}, {Labrie}, {Lim}, {Pierfederici}, {Pontzen}, {Ptak}, {Refsdal}, {Servillat}, \& {Streicher}}]{astropy:2013}
{Astropy Collaboration}, {Robitaille}, T.~P., {Tollerud}, E.~J., {et~al.} 2013, \aap, 558, A33, \dodoi{10.1051/0004-6361/201322068}

\bibitem[{{Astropy Collaboration} {et~al.}(2018){Astropy Collaboration}, {Price-Whelan}, {Sip{\H{o}}cz}, {G{\"u}nther}, {Lim}, {Crawford}, {Conseil}, {Shupe}, {Craig}, {Dencheva}, {Ginsburg}, {Vand erPlas}, {Bradley}, {P{\'e}rez-Su{\'a}rez}, {de Val-Borro}, {Aldcroft}, {Cruz}, {Robitaille}, {Tollerud}, {Ardelean}, {Babej}, {Bach}, {Bachetti}, {Bakanov}, {Bamford}, {Barentsen}, {Barmby}, {Baumbach}, {Berry}, {Biscani}, {Boquien}, {Bostroem}, {Bouma}, {Brammer}, {Bray}, {Breytenbach}, {Buddelmeijer}, {Burke}, {Calderone}, {Cano Rodr{\'\i}guez}, {Cara}, {Cardoso}, {Cheedella}, {Copin}, {Corrales}, {Crichton}, {D'Avella}, {Deil}, {Depagne}, {Dietrich}, {Donath}, {Droettboom}, {Earl}, {Erben}, {Fabbro}, {Ferreira}, {Finethy}, {Fox}, {Garrison}, {Gibbons}, {Goldstein}, {Gommers}, {Greco}, {Greenfield}, {Groener}, {Grollier}, {Hagen}, {Hirst}, {Homeier}, {Horton}, {Hosseinzadeh}, {Hu}, {Hunkeler}, {Ivezi{\'c}}, {Jain}, {Jenness}, {Kanarek}, {Kendrew}, {Kern}, {Kerzendorf}, {Khvalko}, {King}, {Kirkby}, {Kulkarni},
  {Kumar}, {Lee}, {Lenz}, {Littlefair}, {Ma}, {Macleod}, {Mastropietro}, {McCully}, {Montagnac}, {Morris}, {Mueller}, {Mumford}, {Muna}, {Murphy}, {Nelson}, {Nguyen}, {Ninan}, {N{\"o}the}, {Ogaz}, {Oh}, {Parejko}, {Parley}, {Pascual}, {Patil}, {Patil}, {Plunkett}, {Prochaska}, {Rastogi}, {Reddy Janga}, {Sabater}, {Sakurikar}, {Seifert}, {Sherbert}, {Sherwood-Taylor}, {Shih}, {Sick}, {Silbiger}, {Singanamalla}, {Singer}, {Sladen}, {Sooley}, {Sornarajah}, {Streicher}, {Teuben}, {Thomas}, {Tremblay}, {Turner}, {Terr{\'o}n}, {van Kerkwijk}, {de la Vega}, {Watkins}, {Weaver}, {Whitmore}, {Woillez}, {Zabalza}, \& {Astropy Contributors}}]{astropy:2018}
{Astropy Collaboration}, {Price-Whelan}, A.~M., {Sip{\H{o}}cz}, B.~M., {et~al.} 2018, \aj, 156, 123, \dodoi{10.3847/1538-3881/aabc4f}

\bibitem[{{Astropy Collaboration} {et~al.}(2022){Astropy Collaboration}, {Price-Whelan}, {Lim}, {Earl}, {Starkman}, {Bradley}, {Shupe}, {Patil}, {Corrales}, {Brasseur}, {N{"o}the}, {Donath}, {Tollerud}, {Morris}, {Ginsburg}, {Vaher}, {Weaver}, {Tocknell}, {Jamieson}, {van Kerkwijk}, {Robitaille}, {Merry}, {Bachetti}, {G{"u}nther}, {Aldcroft}, {Alvarado-Montes}, {Archibald}, {B{'o}di}, {Bapat}, {Barentsen}, {Baz{'a}n}, {Biswas}, {Boquien}, {Burke}, {Cara}, {Cara}, {Conroy}, {Conseil}, {Craig}, {Cross}, {Cruz}, {D'Eugenio}, {Dencheva}, {Devillepoix}, {Dietrich}, {Eigenbrot}, {Erben}, {Ferreira}, {Foreman-Mackey}, {Fox}, {Freij}, {Garg}, {Geda}, {Glattly}, {Gondhalekar}, {Gordon}, {Grant}, {Greenfield}, {Groener}, {Guest}, {Gurovich}, {Handberg}, {Hart}, {Hatfield-Dodds}, {Homeier}, {Hosseinzadeh}, {Jenness}, {Jones}, {Joseph}, {Kalmbach}, {Karamehmetoglu}, {Ka{l}uszy{'n}ski}, {Kelley}, {Kern}, {Kerzendorf}, {Koch}, {Kulumani}, {Lee}, {Ly}, {Ma}, {MacBride}, {Maljaars}, {Muna}, {Murphy}, {Norman}, {O'Steen},
  {Oman}, {Pacifici}, {Pascual}, {Pascual-Granado}, {Patil}, {Perren}, {Pickering}, {Rastogi}, {Roulston}, {Ryan}, {Rykoff}, {Sabater}, {Sakurikar}, {Salgado}, {Sanghi}, {Saunders}, {Savchenko}, {Schwardt}, {Seifert-Eckert}, {Shih}, {Jain}, {Shukla}, {Sick}, {Simpson}, {Singanamalla}, {Singer}, {Singhal}, {Sinha}, {Sip{H{o}}cz}, {Spitler}, {Stansby}, {Streicher}, {{{S}}umak}, {Swinbank}, {Taranu}, {Tewary}, {Tremblay}, {Val-Borro}, {Van Kooten}, {Vasovi{'c}}, {Verma}, {de Miranda Cardoso}, {Williams}, {Wilson}, {Winkel}, {Wood-Vasey}, {Xue}, {Yoachim}, {Zhang}, {Zonca}, \& {Astropy Project Contributors}}]{astropy:2022}
{Astropy Collaboration}, {Price-Whelan}, A.~M., {Lim}, P.~L., {et~al.} 2022, \apj, 935, 167, \dodoi{10.3847/1538-4357/ac7c74}

\bibitem[{Aublin(2019)}]{Aublin:Antares11Yr}
Aublin, J. 2019, PoS, ICRC2019, 840, \dodoi{10.22323/1.358.0840}

\bibitem[{Braun {et~al.}(2008)Braun, Dumm, De~Palma, Finley, Karle, \& Montaruli}]{Braun:LLH}
Braun, J., Dumm, J., De~Palma, F., {et~al.} 2008, Astropart. Phys., 29, 299, \dodoi{10.1016/j.astropartphys.2008.02.007}

\bibitem[{Cartraud {et~al.}(2023)Cartraud, Aublin, Marinelli, Kouchner, De~La Torre~Luque, Gaggero, Grasso, \& De~Benedittis}]{ANTARES:KRAgTemplate2023}
Cartraud, T., Aublin, J., Marinelli, A., {et~al.} 2023, PoS, ICRC2023, 1084, \dodoi{10.22323/1.444.1084}

\bibitem[{de~la Torre {et~al.}(2022)de~la Torre, Gaggero, Grasso, Fornieri, Evoli, Steppa, \& Egberts}]{Gaggero:KRAgModel}
de~la Torre, P., Gaggero, D., Grasso, D., {et~al.} 2022, {The Galactic diffuse gamma-ray emission meets the PeV frontier - All-sky distributions}, \dodoi{10.5281/zenodo.6385205}

\bibitem[{Deil {et~al.}(2018)}]{GammaCat}
Deil, C., {et~al.} 2018, gamma-cat, https://github.com/gammapy/gamma-cat,  GitHub

\bibitem[{Fuerst {et~al.}(2023)}]{IceCube:2023NTGPICRC}
Fuerst, P.~M., {et~al.} 2023, PoS, ICRC2023, 1046, \dodoi{10.22323/1.444.1046}

\bibitem[{Gaggero {et~al.}(2015)Gaggero, Grasso, Marinelli, Urbano, \& Valli}]{Gaggero:KRAgModelPaper}
Gaggero, D., Grasso, D., Marinelli, A., Urbano, A., \& Valli, M. 2015, Astrophys. J., 815, L25, \dodoi{10.1088/2041-8205/815/2/L25}

\bibitem[{Gaisser {et~al.}(2014)Gaisser, Jero, Karle, \& van Santen}]{Self_Veto:Gaisser}
Gaisser, T.~K., Jero, K., Karle, A., \& van Santen, J. 2014, Phys. Rev., D90, 023009, \dodoi{10.1103/PhysRevD.90.023009}

\bibitem[{{G{\'o}rski} {et~al.}(2005){G{\'o}rski}, {Hivon}, {Banday}, {Wandelt}, {Hansen}, {Reinecke}, \& {Bartelmann}}]{2005ApJ:622:759G}
{G{\'o}rski}, K.~M., {Hivon}, E., {Banday}, A.~J., {et~al.} 2005, ApJ, 622, 759, \dodoi{10.1086/427976}

\bibitem[{Halzen(2006)}]{Halzen:2006AstroPartNu}
Halzen, F. 2006, Eur. Phys. J. C, 46, 669, \dodoi{10.1140/epjc/s2006-02536-4}

\bibitem[{Ke {et~al.}(2017)Ke, Meng, Finley, Wang, Chen, Ma, Ye, \& Liu}]{lightgbm}
Ke, G., Meng, Q., Finley, T., {et~al.} 2017, Advances in neural information processing systems, 30, 3146

\bibitem[{Mancina \& Silva(2020)}]{Mancina:2019ICRC}
Mancina, S., \& Silva, M. 2020, PoS, ICRC2019, 954, \dodoi{10.22323/1.358.0954}

\bibitem[{Osborn {et~al.}(2023)Osborn, Mancina, \& Silva}]{IceCube:ESTReSICRC2023}
Osborn, J., Mancina, S., \& Silva, M. 2023, in {38th International Cosmic Ray Conference}.
\newblock \doarXiv{2308.03856}

\bibitem[{Savina {et~al.}(2023)}]{IceCube:NED2023}
Savina, P., {et~al.} 2023, PoS, ICRC2023, 1010, \dodoi{10.22323/1.444.1010}

\bibitem[{Schönert {et~al.}(2009)Schönert, Gaisser, Resconi, \& Schulz}]{Self_Veto:Schonert}
Schönert, S., Gaisser, T.~K., Resconi, E., \& Schulz, O. 2009, Physical Review D, 79, \dodoi{10.1103/physrevd.79.043009}

\bibitem[{Tanabashi {et~al.}(2018)Tanabashi, Hagiwara, Hikasa, Nakamura, Sumino, Takahashi, Tanaka, Agashe, Aielli, Amsler, Antonelli, Asner, Baer, Banerjee, Barnett, Basaglia, Bauer, Beatty, Belousov, Beringer, Bethke, Bettini, Bichsel, Biebel, Black, Blucher, Buchmuller, Burkert, Bychkov, Cahn, Carena, Ceccucci, Cerri, Chakraborty, Chen, Chivukula, Cowan, Dahl, D'Ambrosio, Damour, de~Florian, de~Gouv\^ea, DeGrand, de~Jong, Dissertori, Dobrescu, D'Onofrio, Doser, Drees, Dreiner, Dwyer, Eerola, Eidelman, Ellis, Erler, Ezhela, Fetscher, Fields, Firestone, Foster, Freitas, Gallagher, Garren, Gerber, Gerbier, Gershon, Gershtein, Gherghetta, Godizov, Goodman, Grab, Gritsan, Grojean, Groom, Gr\"unewald, Gurtu, Gutsche, Haber, Hanhart, Hashimoto, Hayato, Hayes, Hebecker, Heinemeyer, Heltsley, Hern\'andez-Rey, Hisano, H\"ocker, Holder, Holtkamp, Hyodo, Irwin, Johnson, Kado, Karliner, Katz, Klein, Klempt, Kowalewski, Krauss, Kreps, Krusche, Kuyanov, Kwon, Lahav, Laiho, Lesgourgues, Liddle, Ligeti, Lin, Lippmann,
  Liss, Littenberg, Lugovsky, Lugovsky, Lusiani, Makida, Maltoni, Mannel, Manohar, Marciano, Martin, Masoni, Matthews, Mei\ss{}ner, Milstead, Mitchell, M\"onig, Molaro, Moortgat, Moskovic, Murayama, Narain, Nason, Navas, Neubert, Nevski, Nir, Olive, Pagan~Griso, Parsons, Patrignani, Peacock, Pennington, Petcov, Petrov, Pianori, Piepke, Pomarol, Quadt, Rademacker, Raffelt, Ratcliff, Richardson, Ringwald, Roesler, Rolli, Romaniouk, Rosenberg, Rosner, Rybka, Ryutin, Sachrajda, Sakai, Salam, Sarkar, Sauli, Schneider, Scholberg, Schwartz, Scott, Sharma, Sharpe, Shutt, Silari, Sj\"ostrand, Skands, Skwarnicki, Smith, Smoot, Spanier, Spieler, Spiering, Stahl, Stone, Sumiyoshi, Syphers, Terashi, Terning, Thoma, Thorne, Tiator, Titov, Tkachenko, T\"ornqvist, Tovey, Valencia, Van~de Water, Varelas, Venanzoni, Verde, Vincter, Vogel, Vogt, Wakely, Walkowiak, Walter, Wands, Ward, Wascko, Weiglein, Weinberg, Weinberg, White, Wiencke, Willocq, Wohl, Womersley, Woody, Workman, Yao, Zeller, Zenin, Zhu, Zhu, Zimmermann, Zyla,
  Anderson, Fuller, Lugovsky, \& Schaffner}]{PDG}
Tanabashi, M., Hagiwara, K., Hikasa, K., {et~al.} 2018, Phys. Rev. D, 98, 030001, \dodoi{10.1103/PhysRevD.98.030001}

\bibitem[{Wakely \& Horan(2007)}]{Wakely:TeVCat}
Wakely, S.~P., \& Horan, D. 2007, in {30th International Cosmic Ray Conference}, Vol.~3, 1341--1344

\bibitem[{Wakely \& Horan(2014-2022)}]{TeVCat}
Wakely, S.~P., \& Horan, D. 2014-2022, TeVCat, http://tevcat.uchicago.edu/

\bibitem[{Yu {et~al.}(2023)Yu, Kheirandish, Liu, \& Niederhausen}]{IceCube:ESTESSeyferts2023}
Yu, S., Kheirandish, A., Liu, Q., \& Niederhausen, H. 2023, in {38th International Cosmic Ray Conference}.
\newblock \doarXiv{2307.15620}

\bibitem[{Zonca {et~al.}(2019)Zonca, Singer, Lenz, Reinecke, Rosset, Hivon, \& Gorski}]{Zonca2019}
Zonca, A., Singer, L., Lenz, D., {et~al.} 2019, Journal of Open Source Software, 4, 1298, \dodoi{10.21105/joss.01298}

\end{thebibliography}
\bibliographystyle{aasjournal}



\end{document}